\documentclass[12pt]{article}

\usepackage{graphicx,amsmath,natbib,multicol}
\usepackage{amsfonts,amssymb,latexsym,textcomp,txfonts}
\usepackage{epsfig}
\usepackage{fullpage}

\renewcommand{\Pr}{\mathsf{Pr}}

\newcommand{\reals}{\mathbb{R}}

\newcommand{\dd}{\mbox{d}}

\newcommand{\normal}{\mathsf{N}}

\newcommand{\DP}{\mathsf{DP}}
\newcommand{\Dir}{\mathsf{Dir}}
\newcommand{\Gam}{\mathsf{G}}

\newcommand{\bet}{\mathsf{Beta}}

\newcommand{\Mult}{\mathsf{Multinom}}
\newcommand{\SB}{\mathsf{SB}}
\newcommand{\bern}{\mathsf{Bernoulli}}

\title{Sparse covariance estimation in heterogeneous samples}
\author{Abel Rodriguez\footnote{Department of Applied Mathematics and Statistics, University of California, Mailstop SOE2,
Santa Cruz, California 95064, U.S.A. (email: abel@soe.ucsc.edu).}, Alex Lenkoski\footnote{Department of Applied Mathematics, Heidelberg University, Heidelberg, Germany (email: lenkoski@stat.washington.edu).}  and Adrian Dobra\footnote{Departments of Statistics, Biobehavioral Nursing and Health Studies and the Center for Statistics and the Social Sciences, Box 354322, University of Washington, Seattle, WA 98195, U.S.A. (email: adobra@uw.edu).}}
\date{}
\begin{document}
\maketitle
\begin{abstract}
Standard Gaussian graphical models (GGMs) implicitly assume that the conditional independence among variables is common to all observations in the sample.  However, in practice, observations are usually collected form heterogeneous populations where such assumption is not satisfied, leading in turn to nonlinear relationships among variables. To tackle these problems we explore mixtures of GGMs; in particular, we consider both infinite mixture models of GGMs and infinite hidden Markov models with GGM emission distributions.  Such models allow us to divide a heterogeneous population into homogenous groups, with each cluster having its own conditional independence structure.  The main advantage of considering infinite mixtures is that they allow us easily to estimate the number of number of subpopulations in the sample.  As an illustration, we study the trends in exchange rate fluctuations in the pre-Euro era.  This example demonstrates that the models are very flexible while providing extremely interesting interesting insights into real-life applications.\\

{Keywords: Covariance selection; Dirichlet process; Gaussian graphical model; Hidden Markov model, Nonparametric Bayes inference}
\end{abstract}
\section{Introduction}

Problems with small sample sizes and large number of unknown parameters represent one of the most challenging areas of current statistical research.  Graphical models deal with this type of ill-posed problems by enforcing sparsity in the conditional dependence structure among outcomes.  More specifically, given a random vector $X=(X_1,\ldots,X_p) \in \reals^p$, a graphical model for $X$ encodes the conditional independence relationships between its components through a $p$-vertex graph $G$, such that vertex $i$ represents component $X_i$ and the lack of an edge between nodes $i$ and $j$ indicates that variables $i$ and $j$ are conditionally independent.  In particular, Gaussian graphical models (GGMs), also known as covariance selection models \citep{dempster_1972}, have become extremely popular in applications ranging from genetics \citep{WeBlDrHuIsSpZuOlMaNe01,CaRo06} to econometrics and finance \citep{CaWe07,DoEiLe08}.  GGMs assume that the joint distribution of $X$ follows a multivariate Gaussian distribution, and therefore conditional independence among variables can be enforced by setting to zero the appropriate off-diagonal elements of the inverse covariance (precision) matrix. 

One important shortcoming of GGMs is that they implicitly assume a linear relationship between variables.  Copulas have been used in the context graphical models to address nonlinearities.  For example, \cite{BeCo02} decompose the joint distribution of $X$ using pairwise copulas; however, the resulting models are computationally difficult to fit, specially when $p$ grows.  An alternative to copulas is to model non-linearities through mixtures of GGMs.  Countable mixture models explain nonlinearities in the conditional expectations as a consequence of hetherogeneity of the population, and can therefore be interpreted as providing adaptive local linear fits \citep{MuErWe96,RoDuGe08b}.

As a motivation for investigating mixtures of GGMs, consider the analysis of gene expression data.  GGMs have been often used in the context of microarray data, where the graph encoding the conditional dependence structure provides information about expression pathways \citep{DoHaJoNeYaWe04,Fr04,CaRo06}.  The implicit assumptions in these models is that the expression pathways are the same for all individuals/tissues in the sample and that expression levels on different genes are linearly related, which might not be justified if the  underlying population is heterogeneous.  Similarly, when studying the relationship between economic variables such as exchange rates, graphical models allow us to identify groups of countries that form economic blocks and understand how these blocks interact with each other.  However, as trade patterns evolve, we expect that both the block membership and the modes in which countries interact might change, making the constant-graph assumption unrealistic.  In both of these setting, mixtures of GGMs not only provide us with a tool to induce sparsity in heterogeneous samples, but also generate highly interpretable models.

The major challenges in implementing mixtures of GGMs are computational, and relate both to the determination of the underlying graph associated with each component in the mixture and to the estimation of number of components.  Indeed, it is well known that the number of possible partitions for the data grows exponentially, a problem that is compounded when we desire to also estimate the number of components in the mixture and the graphical structure corresponding to each cluster.  Work in finite mixtures of graphical models goes back at least to \cite{ThMeChHe99}, who fixed the number of components in the mixture and developed a search algorithm that uses a modified Cheeseman-Stutz approximation to the marginal likelihood coupled with EM steps to estimate component-specific parameters.  However, to the best of our knowledge, the problem of determining the number of components in mixtures of graphical models has not been properly addressed before.

In this paper, we present the first fully Bayesian approach to inference in nonparametric mixtures and infinite hidden Markov models with Gaussian graphical models as kernel/emission distributions.  Using infinite mixture models provides full support in the space of distributions \citep{Lo84,OnCa04}, and allows us to automatically deal with an unknown number of components/states within a simple computational framework.  The hidden Markov models we discuss allow for the graph encoding the conditional independence structure of the data to change over time, an important feature that has been missing in multivariate time series models employing graphical models \citep{CaWe07,wang_west_2009}.  Since the paper focuses on models that have a P{\`o}lya urn representation, we construct marginal samplers \citep{Ne00} that explicitly integrate out the mean and variance of the individual GGMs.  For problems where the the main interest is either prediction or inference the partition structure and/or the graphical structure associated with each component, this approach greatly reduces computational complexity by avoiding the explicit representation the mean and variance of the different components/states.  Although the paper develops models based on GGMs, the approaches we discuss are not restricted to multivariate continuous outcomes, but can be extended to incorporate combinations of binary, ordinal and continuous variables by introducing latent auxiliary variables.

To simplify our exposition we begin by reviewing Bayesian approaches to inference in Gaussian graphical models in Section \ref{sec:ggmiid} and introducing Dirichlet process mixtures of Gaussian graphical models in Section \ref{sec:ggmddp} and \ref{se:compu}.  We then move to discuss more general nonparametric mixture models in Section \ref{se:gennpggm}, including species sampling mixtures of GGMs and infinite hidden Markov models with GGM emission distributions.  These models are illustrated in Section \ref{sec:illus} with a simulated and a real-world dataset.  Finally, we conclude in Section \ref{se:discussion} with a discussion of possible extensions and future research directions.

\section{Bayesian Framework for  GGMs} \label{sec:ggmiid}

\indent Let $X=X_V$ be the vector of observed variables, where $V=\{1,2,\ldots,p\}$. We assume that $X$ follows a multivariate Normal distribution $p(X|\, \mu,K)= N_p( \mu,K^{-1})$ with mean vector $\mu\in \reals^p$ and $p\times p$ precision matrix $K=(K_{ij})$. We consider the set $\mathcal{G}_V$ of decomposable graphs associated with $V$. The Gaussian graphical model (GGM) associated with a graph $G=(V,E)\in \mathcal{G}_V$ is obtained by setting to zero the elements of $K$ corresponding with missing edges in $G$ \citep{dempster_1972}. The absence of the edge $(i,j)\in (V\times V)\setminus E$ implies $K_{ij}=K_{ji}=0$ which in turn implies that $X_i$ and $X_j$ are conditionally independent given $X_{V\setminus \{ i,j\}}$, i.e.
$$
 X_i\Perp X_j \mid X_{V\setminus \{ i,j\}}.
$$
The precision matrix $K$ belongs to the cone $P_G$ of the symmetric positive definite matrices with entries equal to zero for all $(i,j)\in (V\times V)\setminus E$ \citep{atay-kayis_massam_2005}.  The conditional dependence relationships implied by $G$ induce the following factorization of the joint distribution of $X$ \citep{dawid-lauritzen-1993}:
\begin{align} \label{eq:factlik}
p(X | \, \mu, K,G) &= \frac
{\prod_{C\in \mathcal{C}} p(X_C | \, \mu_{C}, K_{C}) }
{\prod_{S\in \mathcal{S}} p(X_S | \, \mu_{S}, K_{S}) }
\end{align}
where $\mathcal{C}$ denotes the cliques of $G$ and $\mathcal{S}$ denotes separators of $G$. For an index set $V_{0}\subset V$, $\mu_{V_{0}}$  is the subvector of $\mu$ corresponding to the entries in $V_{0}$, while $K_{V_{0}}=((K^{-1})_{V_{0}})^{-1}$. We remark that the subgraph $G_{C}=(C,E_{C})$, $E_{C}=\{(i,j)\in E:i,j\in C\}$, associated with a clique $C\in \mathcal{C}$ is complete, that is, there is no edge missing from it. Similarly, the subgraph $G_{S}$ associated with a separator $S\in \mathcal{S}$ is also complete.

\subsection{Prior specification}

\indent We consider the following joint prior distribution for $\mu$ and $K$:
\begin{eqnarray} \label{eq:jointprior}
 p(\mu,K|G) = p(\mu|K,G)p(K|G),
\end{eqnarray}
where, conditional on $K$, the prior for the mean is $p(\mu|K,G) = N_p(\mu_0,(n_0K)^{-1})$ with $\mu_0\in \reals^p$ and $n_0>0$. The prior for the precision matrix $p(K|G)=W_G(\delta_0,D_0)$ is a G-Wishart distribution with density \citep{roverato_2002,atay-kayis_massam_2005,letac_massam_2007}
\begin{eqnarray} \label{eq:gwishart}
\frac{1}{I_G(\delta_0,D_0)}(\mbox{det}\; K)^{(\delta_0-2)/2}\exp\left\{-\frac{1}{2}\langle K,D_0\rangle\right\},
\end{eqnarray}
with respect to the Lebesgue measure on $P_G$. Here $\langle B,C\rangle = \mbox{tr}(B^{T}C)$ denotes the trace inner product. \citet{diaconis_ylvisaker_1979} prove that the normalizing constant $I_G(\delta_0,D_0)$ is finite if $\delta_0>2$ and $D_0^{-1}\in P_G$. If $G$ is complete (i.e.  $G$ has only one clique $\mathcal{C}=\{V\}$ and no separators), $W_G(\delta_{0},D_{0})$ reduces to the Wishart distribution $W_p(\delta_{0},D_{0})$, hence its normalizing constant is given by
\begin{eqnarray} \label{eq:normconstcomplete}
I_G(\delta_{0},D_{0}) & = & 2^{(\delta_{0}+p-1)p/2} \Gamma_{p}\left\{(\delta_{0}+p-1)/2\right\}(\mbox{det}\; D_{0})^{-(\delta_{0}+p-1)/2},
\end{eqnarray}
\noindent where $\Gamma_{p}(a)=\pi^{p(p-1)/4}\prod_{i=0}^{p-1}\Gamma\left(a-\frac{i}{2}\right)$ for $a>(p-1)/2$ \citep{muirhead_2005}. If $G$ is decomposable but not necessarily complete, \citet{dawid-lauritzen-1993} prove that the G-Wishart distribution $W_G(\delta_{0},D_{0})$ can be factorized according to the cliques and the separators of $G$, hence its normalizing constant is equal to \citep{roverato_2002}:
\begin{eqnarray} \label{eq:normconstreducible}
 I_G(\delta_{0},D_{0}) & = & \frac{\prod_{C\in \mathcal{C}} I_{G_{C}}(\delta_{0},(D_{0})_{C})}{\prod_{S\in \mathcal{S}} I_{G_{S}}(\delta_{0},(D_{0})_{S_j})}. 
\end{eqnarray}
\noindent The subgraphs $G_{C}$ and $G_{S}$ associated with each clique and separator of $G$ are complete, thus $I_{G_{C}}(\delta_{0},(D_{0})_{C})$ and $I_{G_{S}}(\delta_{0},(D_{0})_{S})$ are explicitly calculated as in (\ref{eq:normconstcomplete}). 

\subsection{Posterior distributions and the marginal likelihood of a graph}

The likelihood function for $\mu$ and $K$ corresponding with an $i.i.d.$ sample $x^{(1:n)}=\left( x^{(1)},\ldots,x^{(n)}\right)$  from $X \sim N_p(\mu,K^{-1})$ is given by
\begin{eqnarray} \label{eq:mvnlik}
 L(\mu,K|x^{(1:n)}) = (2\pi)^{-np/2}(\mbox{det}\; K)^{n/2} \exp\left\{ -\frac{1}{2}
 \langle K,U + n(\bar{x}-\mu)(\bar{x}-\mu)^{T}\rangle\right\},
\end{eqnarray}
where $\bar{x}=\frac{1}{n}\sum_{i=1}^nx^{(i)}$, $U=\sum_{i=1}^n(x^{(i)}-\bar{x})(x^{(i)}-\bar{x})^T$. The joint prior (\ref{eq:mvnlik}) is conjugate to the likelihood (\ref{eq:mvnlik}). We assume that the data $x^{(1:n)}$ have been centered and scaled to unit variance, so that the sample mean of each $X_i$ is zero and its sample variance is one. We complete the prior specification by taking $\mu_0=0$, $\delta_0=3$ and $D_0=I_p$, where $I_p$ is the $p$-dimensional identity matrix. The
interpretation of the resulting prior is that the components of $X$ are apriori independent and that the posterior ``weight'' of the prior is equivalent to one observed sample. Other possible choices for the G-Wishart prior parameters $\delta_0$ and $D_0$ are discussed in \citet{carvalho_scott_2009}.\\
\indent The marginal likelihood 
$$p(x^{(1:n)}|G) = \int\limits_{\mu\in \reals^p} \int\limits_{K\in P_G} L(\mu,K|x^{(1:n)}) p(\mu,K|G)\; d K\; d\mu$$
associated with a graph $G\in \mathcal{G}_{V}$ is 
\begin{eqnarray} \label{eq:marglikraw}
 \frac{(2\pi)^{-\frac{(n+1)p}{2}}n_0^{p/2}}{I_G(\delta_0,D_0)}\int\limits_{K\in
   P_G} J(K) (\mbox{det}\;
K)^{\frac{\delta_0+n}{2}}\exp\left\{ -\frac{1}{2} \langle
  K,U+D_0\rangle\right\}\; dK,
\end{eqnarray}
where
\begin{eqnarray*}
 J(K) & = & \int\limits_{\mu\in \reals^p} \exp\left\{ -\frac{1}{2}\langle
   K,
   n(\bar{x}-\mu)(\bar{x}-\mu)^T+n_0(\mu-\mu_0)(\mu-\mu_o)^T\rangle\right\}\;d\mu,\\
 & = & \int\limits_{\mu\in \reals^p} \exp\left\{ -\frac{1}{2}\langle
   K, (n+n_0)( \mu-\bar{\mu})(\mu-\bar{\mu})^T +A\rangle \right\}\;
 d\mu.
\end{eqnarray*}
with $\bar{\mu}=\frac{n\bar{x}+n_0\mu_0}{n+n_0}$ and $A = -(n+n_0)\bar{\mu}\bar{\mu}^T+n \bar{x}\bar{x}^{T} + n_0 \mu_0 \mu_0^T$. After factorizing out $\exp\left\{ -\frac{1}{2} \langle
  K,A\rangle\right\}$, the remaining integrand is the kernel of the posterior distribution of $\mu$:
\begin{eqnarray} \label{eq:muposterior}
 p(\mu|x^{(1:n)},K,G)=N_p\left( \bar{\mu},[(n+n_0)K]^{-1}\right).
\end{eqnarray}

Therefore
\begin{eqnarray*}
 J(K) = \frac{(2\pi)^{p/2}}{(n+n_0)^{p/2}}(\mbox{det}\; K)^{-1/2} \exp\left\{ -\frac{1}{2} \langle
  K,A\rangle\right\}.
\end{eqnarray*}
and it follows that the marginal likelihood
(\ref{eq:marglikraw}) becomes
\begin{eqnarray*}
 \frac{(2\pi)^{-\frac{np}{2}}}{{I_G(\delta_0,D_0)}}\left(
   \frac{n_0}{n+n_0}\right)^{p/2}\int\limits_{K\in P_G} (\mbox{det}\;
 K)^{\frac{\delta_0+n-2}{2}} \exp\left\{ -\frac{1}{2} \langle K, D_0+U+A\rangle\right\}\; dK.
\end{eqnarray*}

The integrand in the equation above is the kernel of the G-Wishart posterior distribution of $K$:
\begin{eqnarray} \label{eq:Kposterior}
 p(K|x^{(1:n)},G)=W_G(\delta_0+n,D_0+U+A).
\end{eqnarray}
and therefore, the final form of the marginal likelihood of the data given $G$ is
\begin{eqnarray} \label{eq:marglik}
 p(x^{(1:n)}|G) = (2\pi)^{-\frac{np}{2}}\left(
   \frac{n_0}{n+n_0}\right)^{p/2} \frac{I_G(\delta_0+n,D_0+U+A)}{I_G(\delta_0,D_0)}.
\end{eqnarray}

A similar argument shows that the posterior predictive distribution of a new sample $x^{(n+1)}$ is given by
\begin{eqnarray}
 p(x^{(n+1)}|x^{(n)},G) =  (2\pi)^{-\frac{p}{2}}\left(
   \frac{n+n_0}{n+1+n_0}\right)^{p/2} \frac{I_G(\delta_0+n+1,D_0+U+A+\widetilde{A})}{I_G(\delta_0+n,D_0+U+A)},\label{eq:preddistrib}
\end{eqnarray}
\noindent where $\widetilde{A} =
-(n+1+n_0)\widetilde{\mu}\widetilde{\mu}^T+ x^{(n+1)}(x^{(n+1)})^T +
(n+n_0) \bar{\mu} \bar{\mu}^T$ and $\widetilde{\mu}=\frac{x^{(n+1)}+(n+n_0)\bar{\mu}}{n+1+n_0}$. Since $G$ is assumed to be decomposable, the posterior normalizing constant $I_G(\delta_0+n,D_0+U+A)$ can be calculated directly using a formula similar to equation (\ref{eq:normconstreducible}), hence $p(x^{(n+1)}|G)$ and $p(x^{(n+1)}|x^{(n)},G)$ can also be calculated directly without any numerical approximation techniques. These computations are key to a successful implementation of the sampling algorithms we describe in Section \ref{se:compu}.

\section{Dirichlet Process Mixtures of GGMs} \label{sec:ggmddp}

Consider now a mixture of GGMs 
\begin{align}\label{eq:finmixt1}
X | \{ w_l\}, \{ \mu_l^{*} \}, \{ K_l^{*} \}, \{ G_l^{*}\}  &\sim \sum_{l=1}^{L} w_l p(X | \, \mu_l^{*}, K_{l}^{*}, G_l^{*})
\end{align}
where $p(X | \, \mu_l^{*}, K_{l}^{*}, G_l^{*})$ is given by \eqref{eq:factlik}.  In this model, draws from from $X$ come from one of $L$ potentially different graphical models; a realization $x^{(i)}$ comes from the $l$-th graphical model (which is defined by the parameters $\mu_l^{*}$, $K_{l}^{*}$ and $G_l^{*}$) independently with probability $w_l$.  A fully Bayesian specification of the model is completed by eliciting prior for the parameters $(\{ w_l\}_{l=1}^{\infty}, \{ \mu_l^{*} \}_{l=1}^{\infty}, \{ K_l^{*} \}_{l=1}^{\infty}, \{ G_l^{*}\}_{l=1}^{\infty})$.  A common choice is to set $w = (w_1,\ldots,w_L) \sim \Dir(w^0)$ and let the component specific parameters $( \mu_l^{*}, K_l^{*},  G_l^{*})$ be i.i.d. samples from some common distribution $M$.

Finite mixtures as the one described above allow for additional flexibility over regular GGMs by allowing a heterogeneous population to be divided into homogenous groups.  However, estimating finite mixture models involves important practical challenges.  For example, in practice we generally do not know how many components are present in the population.  We could allow $L$ to be random and assign a prior distribution to it, but fitting the resulting model involves the use of reversible-jump Markov chain Monte Carlo (RJMCMC) methods \citep{Gr95}, which are notoriously inefficient for high dimensional mixtures.

As an alternative, this section considers Dirichlet process mixtures of GGMs (GGM-DPM).  Note that \eqref{eq:finmixt1} can be alternative written as
\begin{align}\label{eq:finmixt2}
X | H &\sim \int p(X | \, \mu, K, G) H(\dd \mu, \dd K,  \dd G) & H(\cdot) &= \sum_{l=1}^{L} w_l \delta_{(\mu_l^{*}, K_{l}^{*}, G_l^{*})}(\cdot)
\end{align}
where $\delta_{a}(\cdot)$ denotes the degenerate probability measure putting all its mass on $a$.  Therefore, eliciting a prior on $(\{ w_l\}_{l=1}^{\infty}, \{ \mu_l^{*} \}_{l=1}^{\infty}, \{ K_l^{*} \}_{l=1}^{\infty}, \{ G_l^{*}\}_{l=1}^{\infty})$ is equivalent to defining a prior on the discrete probability measure $H$, one such prior is the Dirichlet process  \citep{Fe73,Fe74}.  A random distribution $H$ is said to follow a Dirichlet process (DP) with baseline measure $M$ and precision parameter $\alpha_{0}$, denoted $\DP(\alpha_{0}, M)$, if it has a representation of the form \citep{Se94}
\begin{align}\label{eq:sbDP}
H(\cdot) &\sim \sum_{l=1}^{\infty} w_l \delta_{\theta^{*}_l}(\cdot),
\end{align}
where $\theta^{*}_1, \theta^{*}_2, \ldots$ are independent and identically distributed samples from the {\it baseline measure} $M$ and $w_l = u_l \prod_{s < l} (1 - u_s)$ where $u_1,u_2,\ldots$ is another independent and identically distributed sample where $u_l \sim \bet(1, \alpha_{0})$.  We refer to the joint distribution on $(w_1, w_2, \ldots)$ induced by the above construction as a stick breaking distribution with parameter $\alpha_{0}$, denoted $\SB(\alpha_{0})$.  The DP mixture (DPM) model is recovered from \eqref{eq:finmixt2} when $H \sim \DP(\alpha_0, M)$ for appropriately chosen hyperparameters $\alpha_0$ and $M$.

Consider now an independent and identically distributed sequence $\theta_1,\ldots,\theta_n$ such that $\theta_j | H \sim H$, where $H \sim DP(\alpha_{0}, M)$.  A useful feature of the DP prior is that the joint distribution for $(\theta_1,\ldots,\theta_n)$ obtained after integrating out the random $H$ is given by a sequence of predictive distributions \citep{BlMQ73} where $\theta_{1} \sim M$ and
\begin{align}\label{eq:puDP1}
\theta_{j+1} | \theta_{j},\ldots,\theta_1,\alpha_0 &\sim \sum_{i=1}^{j} \frac{1}{\alpha_{0} + j} \delta_{\theta_i} + \frac{\alpha_{0}}{\alpha_{0} + j} M, & j & > 1.
\end{align}

The presence of ties in the sequence $\theta_1,\ldots,\theta_n$ sometimes makes it convenient to use an alternative representation where $\theta^{*}_1,\ldots,\theta^{*}_L$ denotes the set of $1 \le L \le n$ unique values among $\theta_1,\ldots,\theta_n$ and $\xi_1,\ldots,\xi_n$ is a sequence of indicator variables such that $\theta_j = \theta^{*}_{\xi_j}$.  Under this representation, \eqref{eq:puDP1} implies that $\theta_1^{*},\theta_2^{*},\ldots$ is a sequence of independent and identically distributed samples from $M$, $\xi_1 = 1$ and
\begin{align}\label{eq:puDP2}
\xi_{j+1} | \xi_{j},\ldots,\xi_1,\alpha_0 &\sim \sum_{l=1}^{L^j} \frac{r_{l}^{j}}{\alpha_{0} + j} \delta_{l} + \frac{\alpha_{0}}{\alpha_{0} + j} \delta_{L^j+1}, & j & > 1,
\end{align}
where $L^j =\max_{i \le j} \{ \xi_i \}$ is the number of distinct values among $\theta_1,\ldots,\theta_j$, and $r_l^{j} = \sum_{i=1}^{j} \mathbf{1}_{(\xi_i = l)}$ is the number of samples among the first $l$ with $\xi_j = l$.  Expressions \eqref{eq:puDP1} and \eqref{eq:puDP2} clearly emphasize that, for any finite sample $x^{(1:n)}$, the number of non-empty components $L^n = L$ in a DPM model is a random parameter in the model.  The prior on $L$ implied by the DP \citep{An74} is given by :
\begin{align}\label{eq:priorL}
p(L | \alpha_0, n) &= S(n,L) n! \alpha_0^{L^n} \frac{\Gamma(\alpha_0)}{\alpha_0 + n} & L&=1,\ldots,n,
\end{align}
where $S(\cdot,\cdot)$ denotes the Stirling number of the first kind.  Therefore, the mean number of non-empty components grows with $\alpha_0$, the concentration parameter.

The DP mixture model is intimately connected to the finite mixture model in \eqref{eq:finmixt1}.  Consider a finite mixture with $N$ components such that
\begin{align}
x^{(j)} | \{ \theta^{*}_l \}_{l=1}^{N},\{ \xi_j \}_{j=1}^{n}  & \sim p(x^{(j)} | \theta^{*}_{\xi_j}), & \xi_j | \alpha_0 &\sim \Mult(\alpha_{0}/N,\ldots,\alpha_{0}/N), & \theta^{*}_l & \sim M.
\end{align}
As $N \to \infty$, the predictive distribution for $x^{(1:n)}$ under this model converges to the one obtained from the DP mixture \citep{GrRi01,IsZa02}.

In the GGM-DPM model we explore in this paper we have $\theta = (\mu, K, G)$ and the baseline measure is defined by
\begin{align}\label{eq:baseline}
M &= p(\mu,K|G)p(G),
\end{align}
where $p(\mu,K|G)$ is given by \eqref{eq:jointprior} and $p(G) \propto 1$ is the uniform prior on $\mathcal{G}_V$. Other choices of priors on $\mathcal{G}_V$ that encourage sparsity or have desirable multiple testing properties are discussed in \cite{jones_et_2005,scott_berger_2006,scott_carvalho_2008}.

The GGM-DPM model is a natural extension of the well-known DP mixture of multivariate normals originally presented in \cite{MuErWe96}, but the introduction of the component-specific graphical structure allows us to induce sparsity in the estimation of the precision matrix associated with the mixture components.  The point estimates provided by the GGM-DPM model we just described can be interpreted as providing doubly-regularized estimates of the cluster-specific covariance matrices; one level of regularization arises because of the introduction of the prior distribution on the number of components, which introduces a penalty structure on the number of cluster equal to the logarithm of \eqref{eq:priorL}, while the second level of regularization arises because of the introduction of the prior $p(G)$ on the graph encoding the cluster-specific conditional independence structure.  It is well known that, for high dimensional problems, estimation of the covariance matrices $\{ K^{-1}_l \}_{l=1}^{L}$ can be extremely unstable and that regularized estimators produce improved results; similar approaches to regularization have recently proved effective in both graphical models \citep{Wainwright06high-dimensionalgraphical} and mixture models \citep{FrRa07}.

Although the model just described induces sparsity on the structure of the component specific covariance matrices $\{ K_l \}$, the fact that we are using a mixture of GGMs as the data generating model means that such sparse structure does not translate into conditional independence for the variables involved.  Indeed, even if for a given pair $(i,j)$ we have $(K_l)_{ij} = 0$ for all $l$, $X_i$ and $X_j$ are not conditionally independent under the GGM-DPM model. Therefore, all conditional independence assumptions derived from the graphs $\{ G_l \}$ are valid only conditional on cluster membership.

\section{Computational implementation of GGM mixtures}\label{se:compu}

As with regular GGM models, the posterior distribution arising from the DDP-GGM model is not analytically tractable because of the sheer size of the space of partitions and accompanying graphs.  Therefore, we resort to MCMC algorithms to explore the features of this complicated posterior distribution.  The literature on MCMC samplers for the DPM model has grown extensively in the last 15 years; the approaches can be roughly divided in three large classes:  collapsed (marginal) Gibbs samplers \citep{ME94,EsWe95,Ne00}, which exploit the exchangeability in the data and the P{\'o}lya urn representation in \eqref{eq:puDP1} and \eqref{eq:puDP2} to construct algorithms that avoid explicitly sampling $H$, blocked samplers \citep{IsJa01,RoPa04,Wa07}, which explicitly represent the mixing distribution $H$, and Reversible Jump samplers \citep{JaNe04,JaNe07}.  In this paper we focus attention on marginal samplers such as the ones described in \cite{Ne00} as a natural option that provides some computational advantages.  Indeed, the structure of the baseline measure $M$ in \eqref{eq:baseline} is such that we can easily integrate the means $\{ \mu_l \}$ and precision matrices $\{ K_{l}\}$ out of the model and create a sampler that acts on the space of partitions and graphs directly, which can dramatically reduce the computational burden.

Given an initial state where the data $x^{(1:n)}$ has been divided into $L$ clusters through indicator variables $\xi_1,\ldots,\xi_n$, and where graphs $G_1,\ldots,G_L$ are associated with each of the components, the algorithm proceeds to sample from the joint distribution of $(L,\{\xi_j\}_{j=1}^n,\{ G_l\}_{l=1}^L,\alpha_{0} | x^{(1:n)})$.  As a first stage we update the sequence of indicators $\{\xi_j\}_{j=1}^n$ (and, implicitly, the number of components $L$) by sequentially sampling each $\xi_j$ for $j=1,\ldots,n$ from its full conditional distribution
\begin{align}\label{eq:pusam}
p(\xi_j | \xi^{-j}=\{\xi_{j'}\}_{j'\ne j}, x^{(1:n)}, \{ G_l\}_{l=1}^{L^{-j}}) &\propto \sum_{l=1}^{L^{-j}+1} q_{jl} \delta_{l},
\end{align}
where
\begin{align*}
q_{jl} &\propto \begin{cases}
r^{-j}_{l} p(x^{(j)}| \{ x^{(j')} : j' \ne j, \xi_{j'} = l\},G_l), & l \le L^{-j}, \\
\alpha_{0} p(x^{(j)} | G_{L^{-j}+1}), & l = L^{-j}+1.
\end{cases}
\end{align*}
In the previous expression, $L^{-j}$ is the number of clusters in the sample (excluding observation $x^{(j)}$), $r_{l}^{-j} = \sum_{j' \ne j} \mathbf{1}_{\{\xi_j'=l\}}$ is the number of observations included in cluster $l$ (excluding observation $j$ if this sample currently belongs to cluster $l$), $p(x^{(j)}| \{ x^{(j')} : j' \ne j, \xi_{j'} = l\},G_l)$ is the posterior predictive distribution of sample $x^{(j)}$ given the samples that are currently in the $l$-th cluster (excluding $x^{(j)}$ if it happens to belong to this cluster) and the graph $G_l$ associated with this cluster \--- see equation \eqref{eq:preddistrib}, and $p(x^{(j)} | G_{L^{-j}+1})$ is the posterior predictive distribution of sample $x^{(j)}$ given an empty cluster, which is calculated by setting $n=0$, $\bar{\mu}=0_{(p\times 1)}$ and $U=0_{(p\times p)}$ in  equation \eqref{eq:preddistrib}.  The graph $G_{L^{-j}+1}$ is to be randomly sampled from our baseline measure on $\mathcal{G}_V$, which we labeled $p(G)$ in \eqref{eq:baseline}.  If the last observations has been moved out of a cluster, that cluster is deleted and $L$ is decreased by $1$. Similarly, if an observation is moved to a new cluster that is currently empty, $L$ is increased by $1$.

Once the cluster assignment has been updated, the graph $G_{l}$ associated with each cluster $l=1,\ldots,L$ is also updated as follows. We let the neighborhood of $G_{l}$, denoted by $\mbox{nbd}_{\mathcal{G}_V}(G_l)$, be the set of decomposable graphs that can be obtained from $G_{l}$ by adding or deleting one edge. These neighborhood sets connect any two graphs in $\mathcal{G}_{V}$ through a sequence of graphs that differ by exactly one edge \--- see, for example, \citet{lauritzen_1996}. We draw a candidate graph $G_l^{new}$ from the uniform distribution on $\mbox{nbd}_{\mathcal{G}_V}(G_l)$. We change the graph associated with cluster $l$ to $G_l^{new}$ with probability
\begin{eqnarray*}
\min\left\{ 1, \frac{p(\{ x^{(j)}:\xi_j=l\}|G_l^{new})/|\mbox{nbd}_{\mathcal{G}_V}(G^{new}_l)|}{p(\{ x^{(j)}:\xi_j=l\}|G_l)/|\mbox{nbd}_{\mathcal{G}_V}(G_l)|}\right\},
\end{eqnarray*}
\noindent otherwise the graph associated with cluster $l$ remains unchanged.
Here $p(\{ x^{(j)}:\xi_j=l\}|G)$ represents the marginal likelihood of the samples currently in cluster $l$ given a graph $G$ \--- see equation \eqref{eq:marglik}.  We denote by $|B|$ the number of elements of a set $B$. To improve mixing, we update the graphs associated with each cluster multiple times before another cluster assignment update is carried out (typically, between 5 and 10 times seems to provide adequate mixing).

These two sequences of steps produce a sample from the posterior distribution of interest, $(L,\{\xi_j\}_{j=1}^n,\{ G_l\}_{l=1}^L,\alpha_{0} | x^{(1:n)})$ without any need to sample the means $\{ \mu \}_{l=1}^{L}$ or precisions $\{ K \}_{l=1}^{L}$.  Therefore, if we are only interested in inferences about the clustering structure or the graphical structure associated with the clusters, or on predictive inference, the previous algorithm is sufficient and can dramatically reduce the computational burden of the algorithm.  However, if needed, the mean and variances of each mixture component can be easily sampled conditional on $\{ \xi_j\}_{j=1}^{n}$ by noting that for every $l=1,\ldots,L$ (see equations \eqref{eq:muposterior} and \eqref{eq:Kposterior})
\begin{align*}
K_l | G_l, \{ x^{(j)} : \xi_j = l \} &\sim W_{G_l} ( \delta_0 + r_l , D_0 + U_l + A_l ), \\
\mu_l | K_l , G_l, \{ x^{(j)} : \xi_j = l \} &\sim \normal_p( \bar{\mu}_l, [(r_l+n_0)K_l]^{-1} ),
\end{align*}
independently of other components.  The subscript $l$ denotes the corresponding values computed using only the observations assigned to component $l$ (for example, $r_l$ is the number of observations assigned to component $l$).

Also, additional flexibility can be obtained by sampling some of the hyperparameters associated with the DP prior.  For example, the concentration parameter $\alpha_0$ controls the expected number of components, and therefore has an important effect on the inferences generated by the model.  Since eliciting values for $\alpha_0$ can be extremely difficult in practice, it is recommendable to try to infer it from the data.  For example, we can assume a vague $\Gam(a_{0},b_{0})$ prior for the precision parameter $\alpha_{0}$, in which case the full conditional distribution can be easily sampled using an auxiliary-variable Gibbs sampling step \citep{EsWe95} (see Appendix).

\section{More general nonparametric mixtures of graphical models}\label{se:gennpggm}

The ideas just described for Dirichlet process mixtures of Gaussian graphical models can be directly extended to other nonparametric mixture models where P{\`o}lya urn representations similar to \eqref{eq:puDP1} and \eqref{eq:puDP2} can be exploited to analytically integrate out  the random distributions out of the model.  Some example include species sampling models \citep{MC65,Pi96,LiMePr07,LeMuTrQu09}, hierarchical Dirichlet processes \citep{TeJoBeBl04}, nested Dirichlet processes \citep{RoDuGe08a} and linear combinations of Dirichlet process \citep{MuQuRo04,DuPi04}.  In this section we consider in detail two such extensions.

\subsection{Species sampling models}

As a first example, consider replacing the Dirichlet process mixture of GGMs with a more general species sampling mixture of GGMs.  An {\it exchangeable} sequence $\theta_1,\ldots,\theta_n$ is said to follow a species sampling model \citep{MC65,Pi96,LiMePr07,LeMuTrQu09} with baseline measure $M$ if $\theta_j = \theta^{*}_{\xi_j}$ where $\theta^{*}_1,\theta^{*}_2,\ldots$ is a sequence of independent and identically distributed samples from $M$ and the indicators $\xi_1,\ldots,\xi_n$ are sampled according to the predictive formula 
\begin{align}\label{eq:predSSM}
\xi_{j+1} | \xi_{j},\ldots,\xi_{1} \sim \sum_{l=1}^{L^j} v_{jl}(r_{1}^{j},\ldots, r_{L^j}^{j}) \delta_{l} + v_{j,L^j + 1} \delta_{L^j},
\end{align}
where the weights $v_{j1},\ldots v_{j,L^j+1}$ satisfy $\sum_{i=1}^{L^j+1} v_{jl} =1$ for all $i$ and $r_{l}^{j}$ and $L^j$ are defined as in Section \ref{sec:ggmddp}.  The Dirichlet process is the best known member of the class of species sampling models, which also includes the two parameter Poisson-Dirichlet process \citep{Pi96,IsJa01} and the normalized inverse-gamma priors\citep{LiMePr05b}, among others.

Moving beyond DPM models is of interest because the prior on the partition structure induced by the Dirichlet process can be somewhat restrictive.  For example, \eqref{eq:priorL} implies that, a priori and for a given precision parameter $\alpha_0$, the expected number of occupied clusters grows with the logarithm of $n$, which might be inappropriate for certain application such as computer vision \citep{SuJo09}.  Also, the Dirichlet process favors partitions that consist of a small number of clusters with a large number of observations along with a larger number of small clusters.

Since samples from a species sampling model are exchangeable, the predictive distribution \eqref{eq:predSSM} also provides the full conditional distribution required to implement the Gibbs sampling algorithm discussed in Section \ref{sec:ggmddp}.  For example, for the Poisson Dirichlet process with baseline measure $M$, discount $0 \le \alpha < 1$ and strength $\alpha_{0} \ge -\alpha$ we have 
\begin{align*}
\xi_{j+1} | \xi_{j},\ldots,\xi_{1} \sim \sum_{l=1}^{L^j} \frac{ r^{j}_{l} - \alpha}{\alpha_{0} + j} \delta_{l} + \frac{\alpha_{0} - \alpha L^j}{\alpha_{0} + j} \delta_{L^j+1}
\end{align*}

Note that if $\alpha = 0$, the Poisson-Dirichlet process reduces to the standard Dirichlet process.  Modifying the algorithm in Section \ref{sec:ggmddp} to fit a Poisson-Dirichlet mixture of GGMs is straightforward. In particular, we only need to slightly modify the posterior weights in \eqref{eq:pusam} to reflect the new prior distribution,
$$
q_{jl} =\begin{cases}
(r_{l}^{-j} - \alpha) p(x^{(j)}| \{ x^{(j')} : j' \ne j, \xi_{j'} = l\},G_l), & l \le L^{-j}, \\
(\alpha_{0} - \alpha L^{-j}) p(x^{(j)} | G_{L^{-j}+1}), & l = L^{-j}+1.
\end{cases}
$$

\subsection{Infinite Hidden Markov Gaussian Graphical models}\label{se:ihmm-ggm}

Recently, multivariate time series models that use graphical models to improve estimation of the crosssectional covariance structure have been developed \citep{CaWe07,wang_west_2009}.  These approaches rely on extensions of the dynamic linear model (DLM) \citep{WeHa97} and assume that the graph underlying the model is constant in time which, as our first illustration in Section \ref{sec:macroecon} demonstrates, might not be an appropriate assumption in practical application.  As an alternative we focus on a nonparametric version of the popular hidden Markov model where the emission distribution corresponds to a GGM.

Hidden Markov models (HMMs) \citep{CaMoRy05}, are hierarchical mixture models where
\begin{align*}
x^{(j)} | \{ \theta^{*}_l \}_{l=1}^{L}, \{ \xi_j \}_{j=n}^{L} &\sim p(x^{(j)} | \theta^{*}_{\xi_j}),  & \xi_j | \xi_{j-1}, \{ \pi^{l}\}_{l-1}^{L} &\sim \Mult( \pi^{\xi_{j-1}} ), & \xi_0& \sim \Mult( \pi^0 ) & \theta^{*}_l&\sim M. 
\end{align*}
In this context the latent indicator $\xi_j\in \{1,\ldots,L\}$ is called a hidden state, while the entire set of indicators $\{ \xi_j\}_{j=1}^n$ is called a trajectory. The ordering of the states is implicitly defined by the ordering of their indices; trajectories evolve according to a Markov process with transition probabilities $\Pr( \xi_j = l | \xi_{j-1} = l')  = \pi^{l'}_{l}$.  The initial state probabilities are $\Pr(\xi_0=l)=\pi^0_l$. Conditionally on a set of states $\{ \xi_j\}_{j=1}^n$,  the observations $x^{(1)},\ldots, x^{(n)}$, are independently distributed from state dependent distributions $p(\cdot | \theta^{*}_{\xi_1}),\ldots,p(\cdot | \theta^{*}_{\xi_n})$. 

Infinite hidden Markov models (iHMMs) \citep{BeGhRa01,TeJoBeBl04,GaSaTeGh08} generalize HMMs to models with an infinite number of states, in a similar way as how Dirichlet process models generalize finite mixture models, allowing us to estimate the number of states $L$.  In particular, we can build a GGM-iHMM where
\begin{align*}
x^{(j)} | \{ \mu_l \}_{l=1}^{\infty},\{ K_l \}_{l=1}^{\infty},\{ G_l \}_{l=1}^{\infty}, \{ \xi_j \}_{j=1}^{n} &\sim \normal_p (x^{(j)} | \mu_{\xi_j}, K^{-1}_{\xi_j}),   \\ 
\xi_j | \xi_{j-1}, \{ \pi^{l} \}_{l=1}^{\infty} &\sim \Mult( \pi^{\xi_{j-1}} ) \\
 \pi^{l} | \alpha, \gamma & \sim \DP(\alpha, \gamma)\\
  \gamma | \alpha_0 &\sim \SB(\alpha_0),
\end{align*}
and $\theta^{*}_l=(\mu_l,K_l,G_l)  \sim M$, where $M$ is defined as in \eqref{eq:baseline}.  This  GGM-iHMM has some distinct advantages over the dynamic linear models with graphical structure discussed in \cite{CaWe07} and \cite{wang_west_2009}.  In particular, it allows for the graph controlling the conditional independence structure of the data to evolve in time while still taking into account the sequential nature of the problem.

Again, a marginal Gibbs sampler similar to the one described in Section \ref{sec:ggmddp} can be devised for the GGM-iHMM.  We denote by $r^{j_1:j_2}_{ll'}$ the number of transitions from state $l$ to state $l'$ in the sub-trajectory $\{ \xi_j\}_{j=j_1}^{j_2}$ and by $r^{j_1:j_2}_{l\cdot}$ the number of transitions out of state $l$ in the same sub-trajectory. Given the base DP parameters $\gamma = \{\gamma_1,\ldots,\gamma_{L+1}\}$ and the precision parameters $\alpha_0$ and $\alpha$, the states $\xi_1,\ldots,\xi_n$ are sequentially updated using the full conditional distributions:
\begin{align*}
q_{jl} = \begin{cases}
\left(r^{1:(j-1)}_{\xi_{j-1},l}+r^{(j+1):n}_{\xi_{j-1},l} + \alpha \gamma_l \right)\frac{ r_{l,\xi_{j+1}}^{1:(j-1)}+r_{l,\xi_{j+1}}^{(j+1):n} + \alpha \gamma_{\xi_{j+1}}}{ r_{l\cdot}^{1:(j-1)}+r_{l\cdot}^{(j+1):n} + \alpha} p(x^{(j)}| \{ x^{(j')} : j' \ne j, \xi_{j'} = l\},G_l),& l \le L^{-j}, l \ne \xi_{t-1}, \\
\left(r^{1:(j-1)}_{\xi_{j-1},l}+r^{(j+1):n}_{\xi_{j-1},l} + \alpha \gamma_l \right)\frac{ r_{l,\xi_{j+1}}^{1:(j-1)}+r_{l,\xi_{j+1}}^{(j+1):n} + \alpha \gamma_{\xi_{j+1}} + 1}{ r_{l\cdot}^{1:(j-1)}+r_{l\cdot}^{(j+1):n} + \alpha + 1} p(x^{(j)}| \{ x^{(j')} : j' \ne j, \xi_{j'} = l\},G_l), & l = \xi_{j-1} = \xi_{j+1}, \\
\left(r^{1:(j-1)}_{\xi_{j-1},l}+r^{(j+1):n}_{\xi_{j-1},l} + \alpha \gamma_l \right)\frac{ r_{l,\xi_{j+1}}^{1:(j-1)}+r_{l,\xi_{j+1}}^{(j+1):n} + \alpha \gamma_{\xi_{j+1}}}{ r_{l\cdot}^{1:(j-1)}+r_{l\cdot}^{(j+1):n} + \alpha + 1} p(x^{(j)}| \{ x^{(j')} : j' \ne j, \xi_{j'} = l\},G_l), & l = \xi_{j-1} \ne \xi_{j+1}, \\
\alpha \gamma_l \gamma_{\xi_{j+1}} p(x^{(j)}| G_{L+1}), & l = L^{-j}+1.
\end{cases}
\end{align*}
\noindent If a new empty cluster needs to be created, we update the number of clusters $L$ by setting $L^{new} = L+1$ and the vector $\gamma$ by setting  $\gamma^{new}_{L+1}=v \gamma_{L+1}$, $\gamma^{new}_{L+2}=(1-v)\gamma_{L+1}$ were $v\sim \mbox{Beta}(\alpha_{0},1)$. \\
\indent Given a trajectory $\{ \xi_j\}_{j=1}^n$, $\alpha_0$ and $\alpha$, we sample $\gamma$ by introducing the independent auxiliary variables $\{ m_{ll'} \}$ for $l, l' \in \{1,\ldots, L\}$ such that 
\begin{align*}
\Pr(m_{ll'} = m) &\propto S(r^{(1:n)}_{ll'}, m) (\alpha \gamma_{l'})^m,  & m &\in\{ 1, \ldots, r^{(1:n)}_{ll'} \},
\end{align*}
where $S(\cdot,\cdot)$ denotes the Stirling number of the first kind. Conditional on these auxiliary variables we can update $\gamma$ by sampling
$$
(\gamma_1,\ldots,\gamma_{L+1}) \sim \Dir(m_{\cdot 1},\ldots,m_{\cdot L},\alpha_0),
$$
where $m_{\cdot l'} =\sum_{l=1}^{L} m_{ll'}$. We use vague gamma priors for the precision parameters $\alpha_0$ and $\alpha$ and update them as described in the Appendix.

\section{Examples}\label{sec:illus}

\subsection{Simulated data}

\begin{figure}
\begin{center}
\includegraphics[width = 3in]{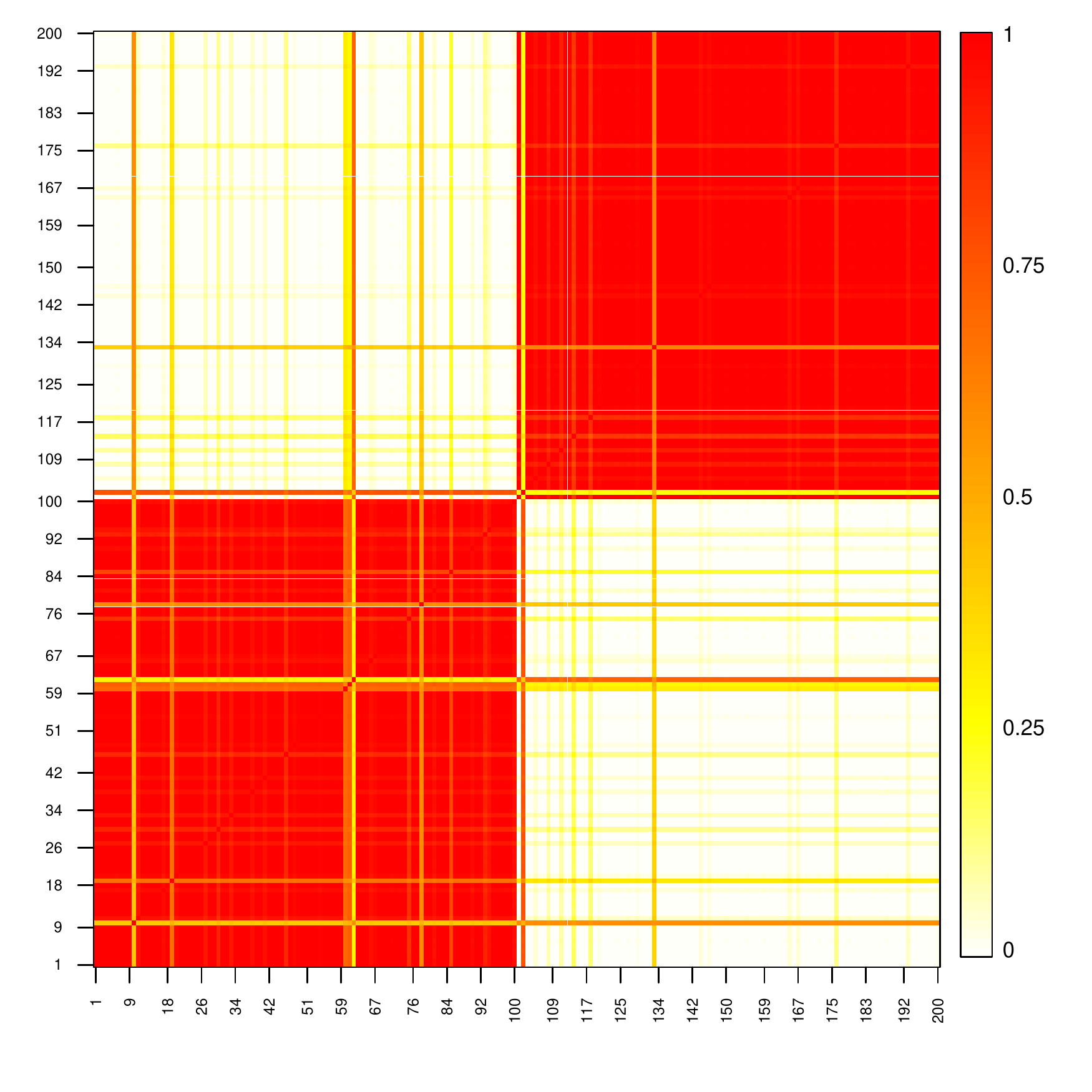}
\includegraphics[width = 3in]{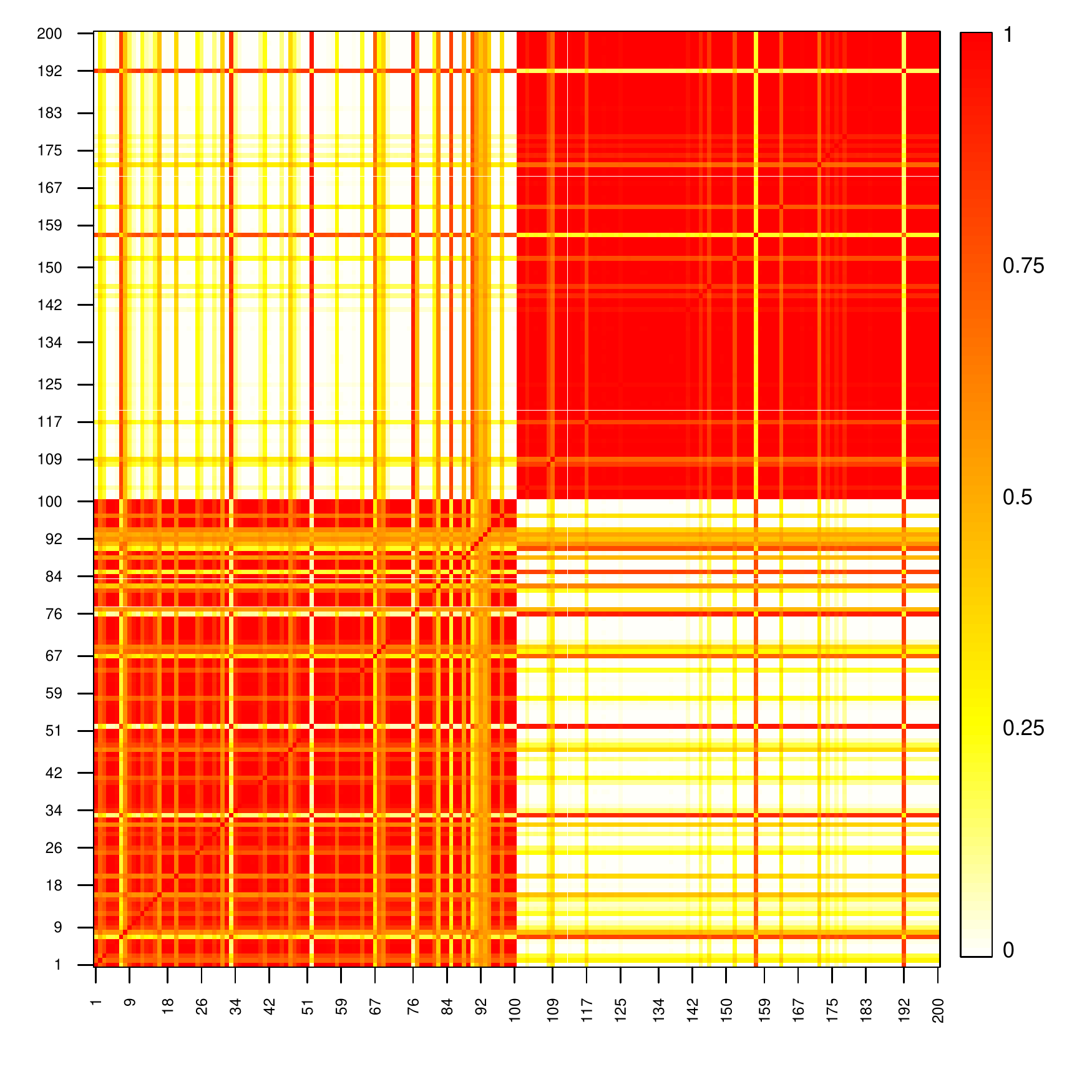}
\includegraphics[width = 3in]{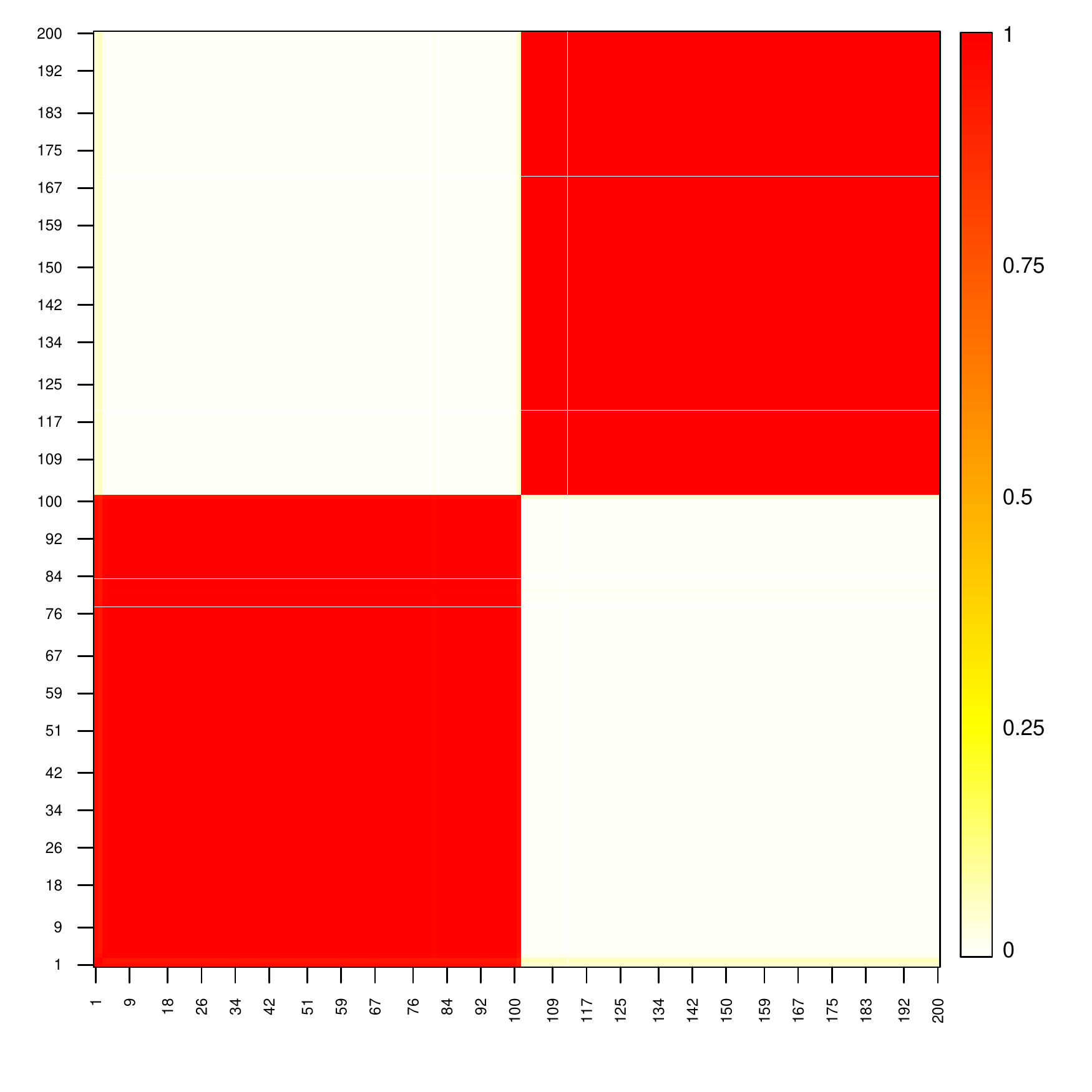}
\includegraphics[width = 3in]{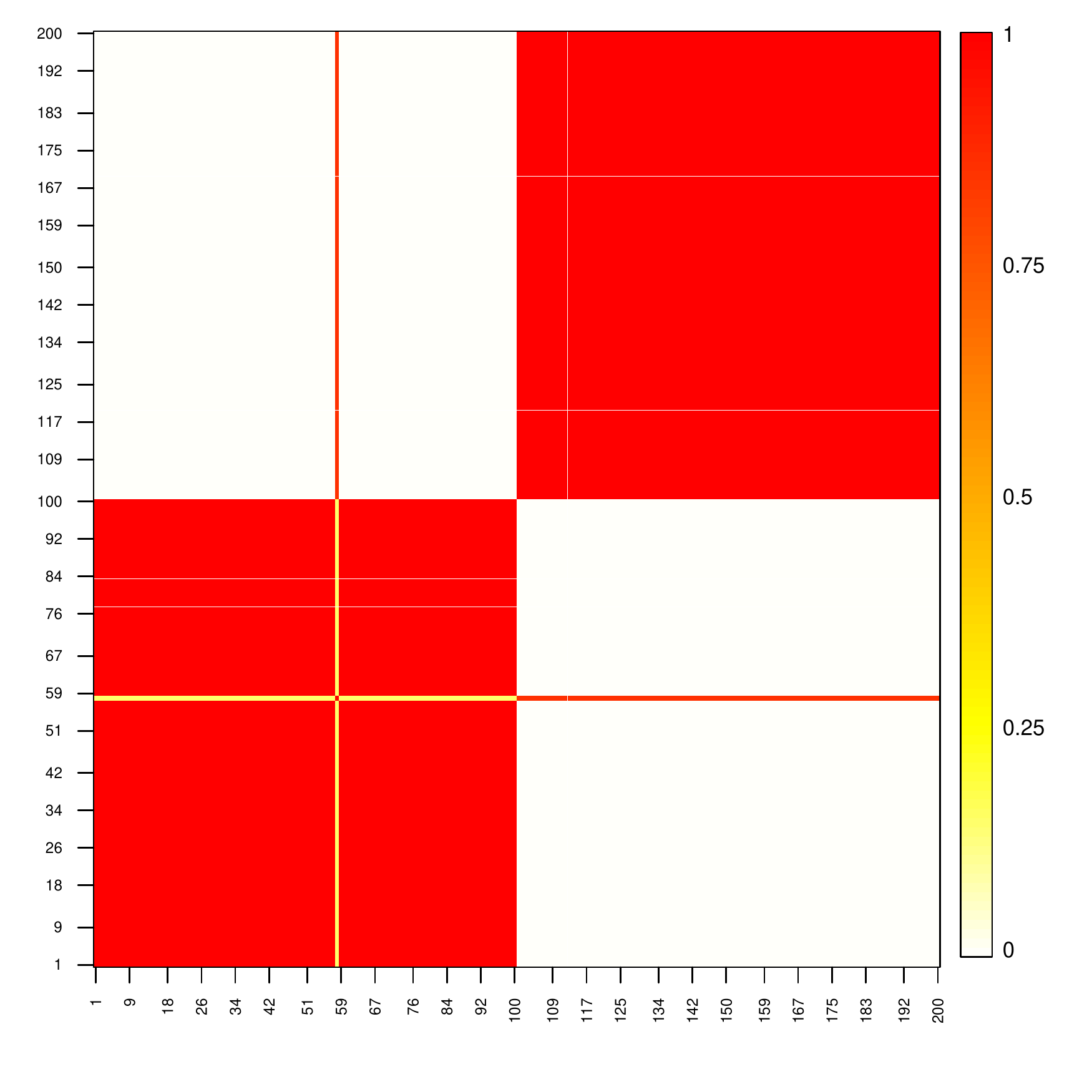}
\end{center}
\caption{Probability that two observations belong to the same cluster in the simulated example.  The panel shows, in a clockwise order, the results using the DPM model in the unrestricted case, the DPM model restricting to the full graph, the iHMM model restricting to the full model and the iHMM model in the unrestricted case.  Note that the true underlying clustering is shown by the iHMM model in the unrestricted case.}\label{fig:sim_results_cluster}
\end{figure}

\begin{figure}
\begin{center}
\includegraphics[width = 2in]{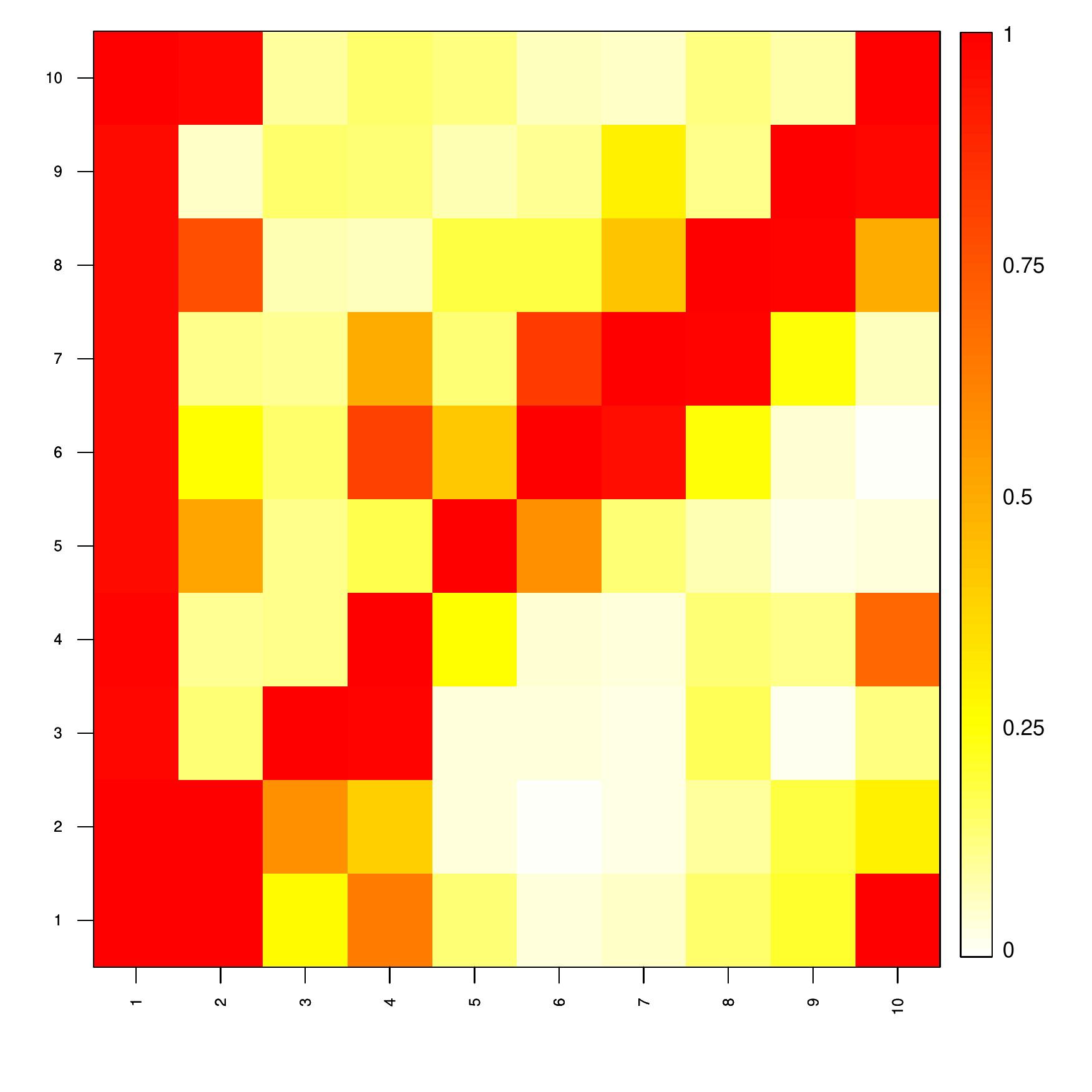}
\includegraphics[width = 2in]{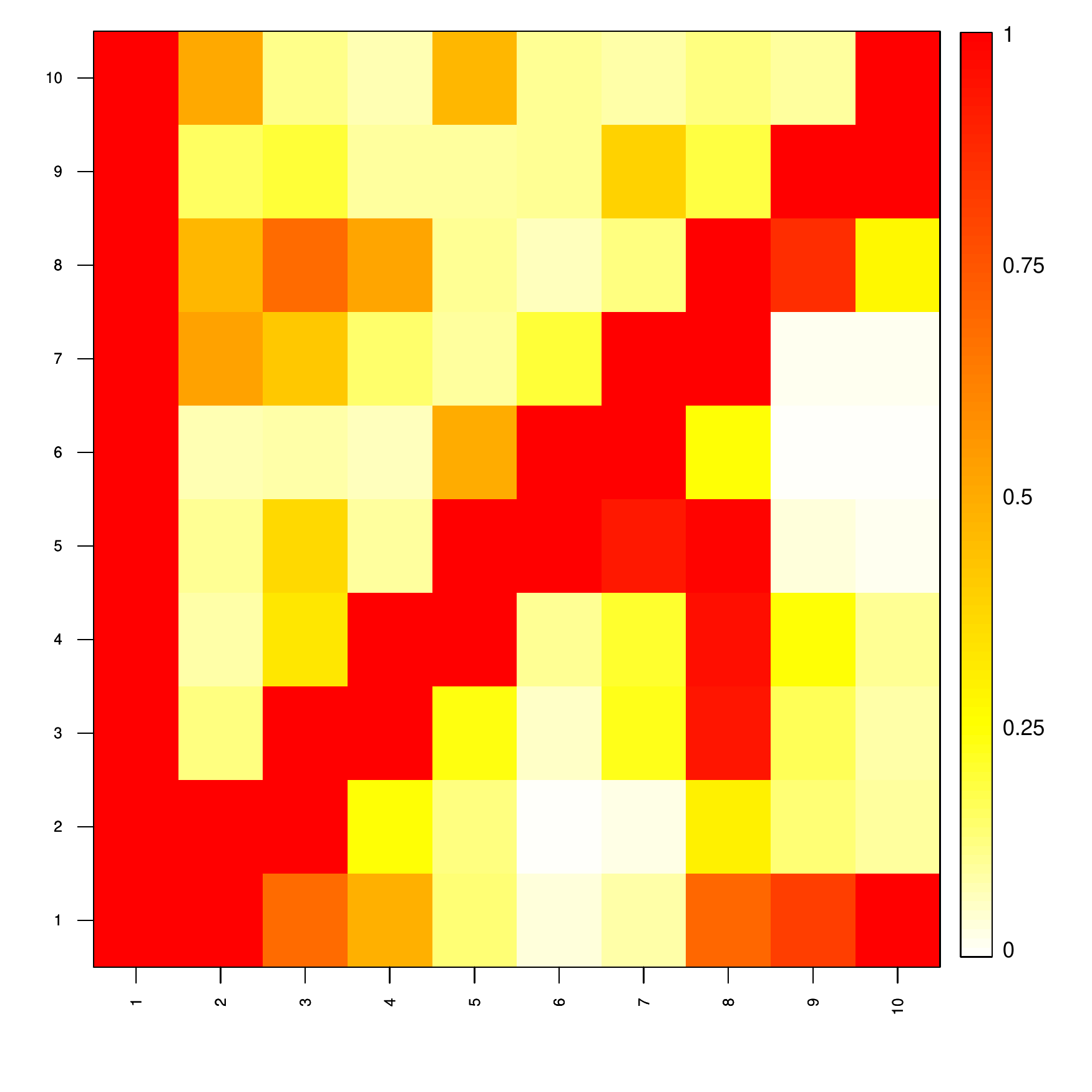}
\includegraphics[width = 2in]{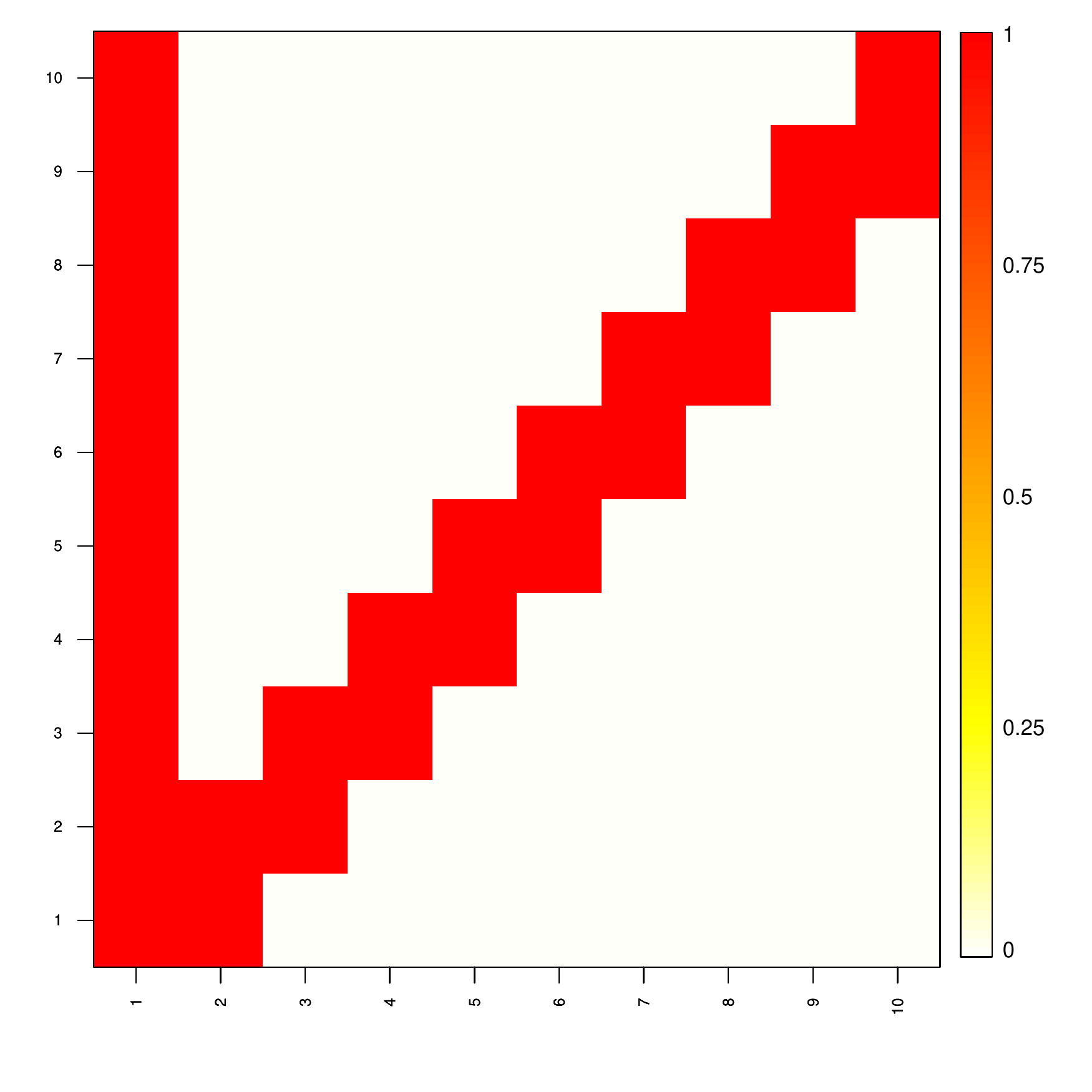}
\end{center}
\caption{Edge probabilities in the simulated example.  In each panel, the upper triangle shows the average edge probability of observations in the first cluster, while the lower triangle shows the average edge probability of observations in the second cluster.  The lefthand panel corresponds to the results from the DPM search, the middle panel corresponds to the iHMM model and the righthand panel displays the true edges in the underlying model.}\label{fig:sim_results_edge}
\end{figure}

We consider first a small simulation that involves data arising from a two Gaussian clusters. For brevity, the results we present in this Section correspond to a single run of the simulation, but these are representative of those obtained over multiple runs.  In the first cluster samples are from a star graphical model $N_{10}(0.5,K_{1}^{-1})$ with every variable $X_{j}$, $j>2$, connected to $X_{1}$, while the second cluster contains $100$ samples from a cycle model $N_{10}(-0.5,K_{2}^{-1})$. The non-zero elements of the two precision matrices are
\begin{align*}
 & (K_{1})_{j,j}=(K_{2})_{j,j}=1, & j =1,\ldots,10,\\
 & (K_{1})_{1,j}=(K_{1})_{j,1} = 0.3, & j=2,\ldots,10,\\
 & (K_{2})_{j-1,j}=(K_{2})_{j,j-1}=0.3, & j=2,\ldots,10,\\
 & (K_{2})_{1,10}=(K_{2})_{10,1}=0.3.&
\end{align*}
We are interested in recovering the two clusters and their corresponding conditional independence graphs.  Since the first $100$ observations belong to the first cluster and the remainder to the second cluster, we are also able to employ the iHMM discussed above and compare its performance to the DPM model.  After sampling a dataset, we ran both DPM and iHMM samplers for 20,000 iterations after 5,000 iterations of burn-in. To determine the utility of including GGMs in the Dirichlet Process framework we also ran both the DPM and iHMM models with attention restricted to the full graph. \\
\indent Figure~\ref{fig:sim_results_cluster} shows the clustering for the DPM and iHMM both using the full space of decomposable GGMs and with attention restricted to the full graph.  The upper lefthand panel, which corresponds to the results from the DPM model using the full graph space, shows two well defined groups with a few observations incorrectly classfied.  The upper righthand panel shows that restricting attention to the full graph decreases the ability to cluster observations properly.  The bottom row of Figure~\ref{fig:sim_results_cluster} shows that using the iHMM model in this case dramatically improves clustering.  The iHMM model using the unrestricted graph space clusters the observations perfectly, while the iHMM model using the full graph does nearly as well, though it misclassifies one observation.\\
\indent Figure~\ref{fig:sim_results_edge} shows the edge probabilities in each of the two clusters for the DPM and iHMM models, along with the true edges from the underlying model.  We see that for the most part, the edges with the highest probability in both models correspond to the true edges.  However, the improved clustering capabilities of the iHMM model translates into sharper edge probabilities, especially in the second cluster.\\

\subsection{Real data: understanding trends in exchange rate fluctuations}\label{sec:macroecon}

\begin{figure}
\centering\includegraphics[width = 3in]{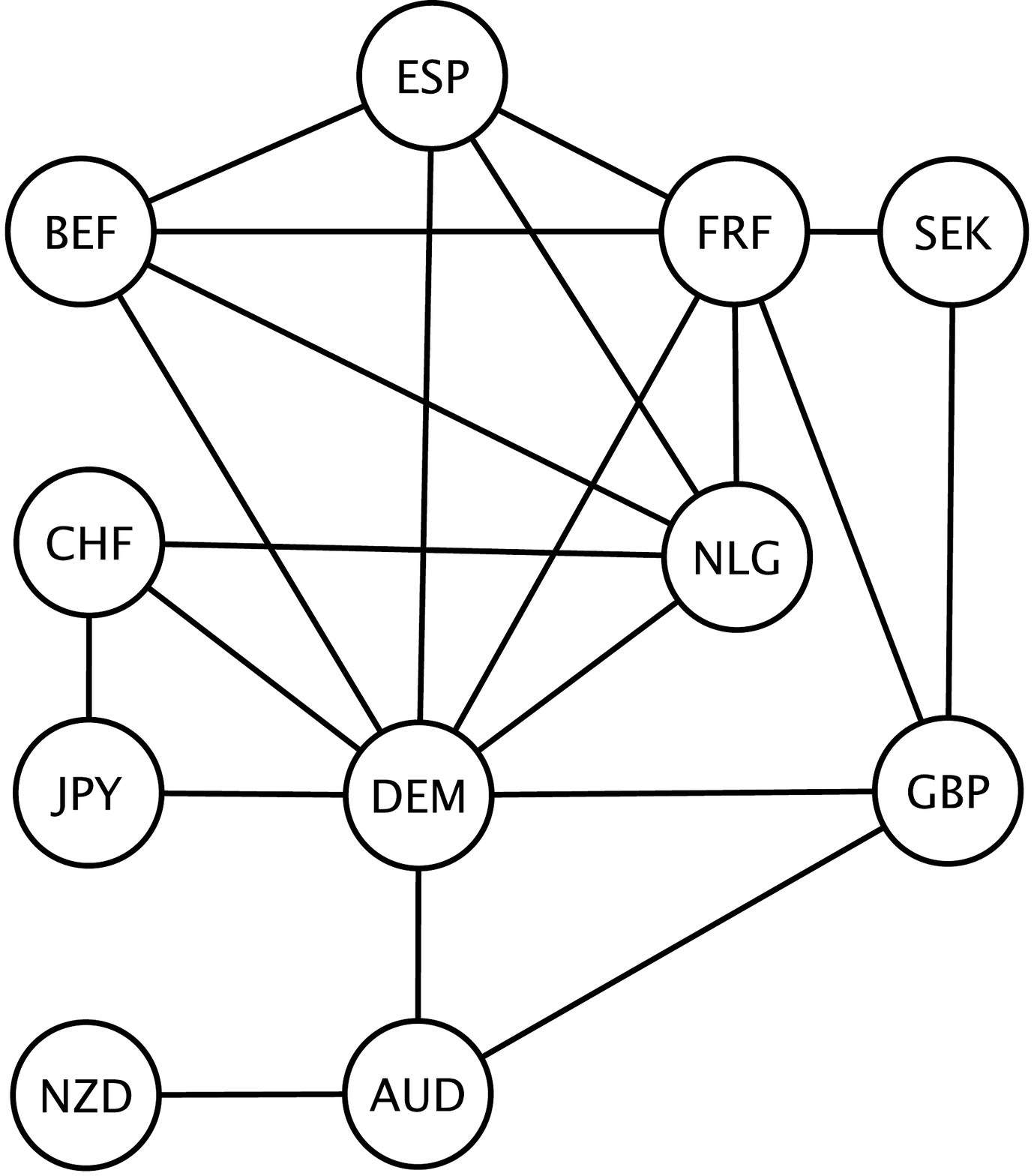}
\caption{Graphical model presented by Carvalho et al \cite{carvalho_et_2007}, which represents the highest probability graph found using stochastic search methods over the last 100 timepoints of the exchange dataset, using the author's prior specifications.  This graph has been used to show that investment strategies based on graphical models often have lower variability and higher yield than methods based on the full covariance matrix.}\label{fig:carvalho_graph}
\end{figure}

\begin{figure}
\centering\includegraphics[width = 2.5in]{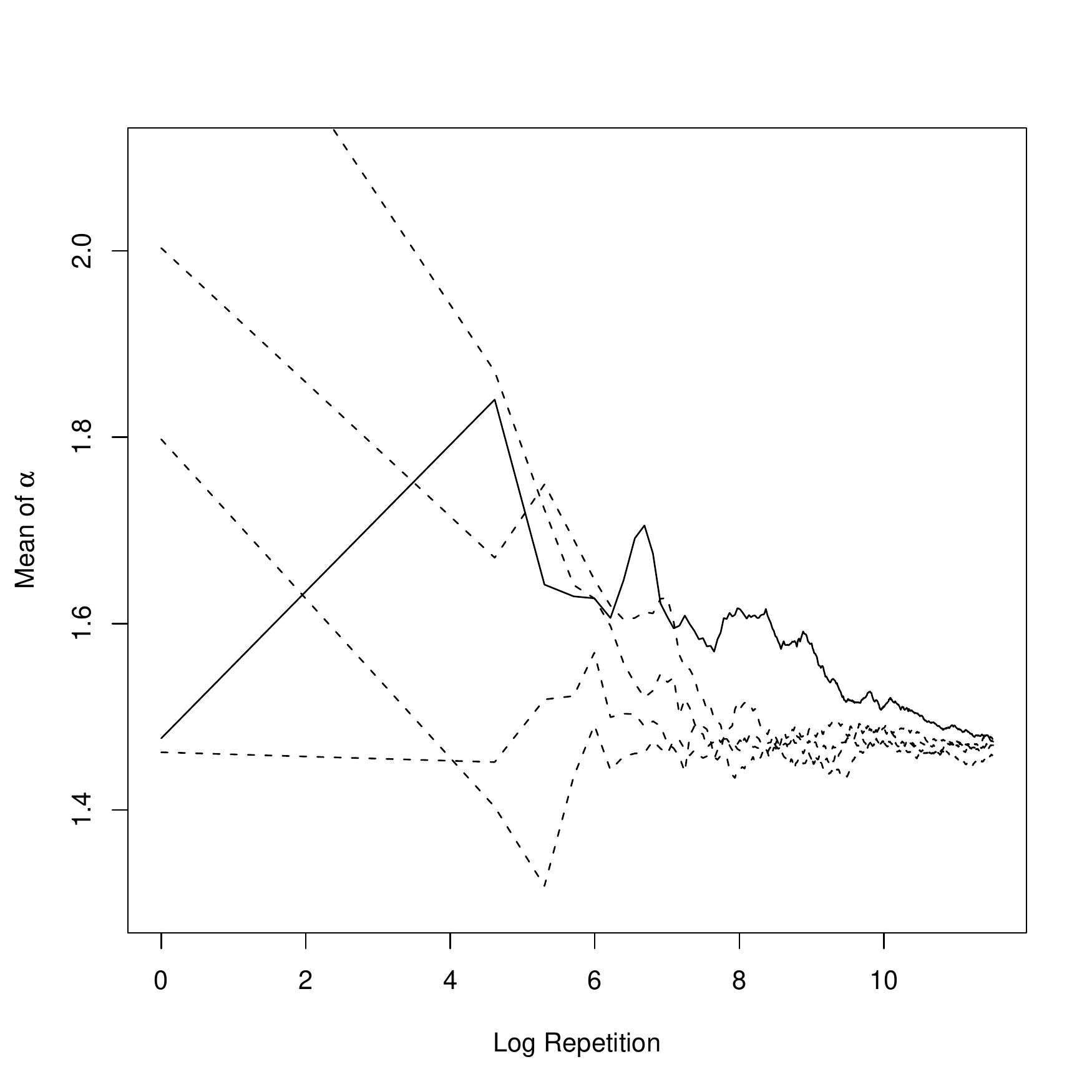}
\centering\includegraphics[width = 2.5in]{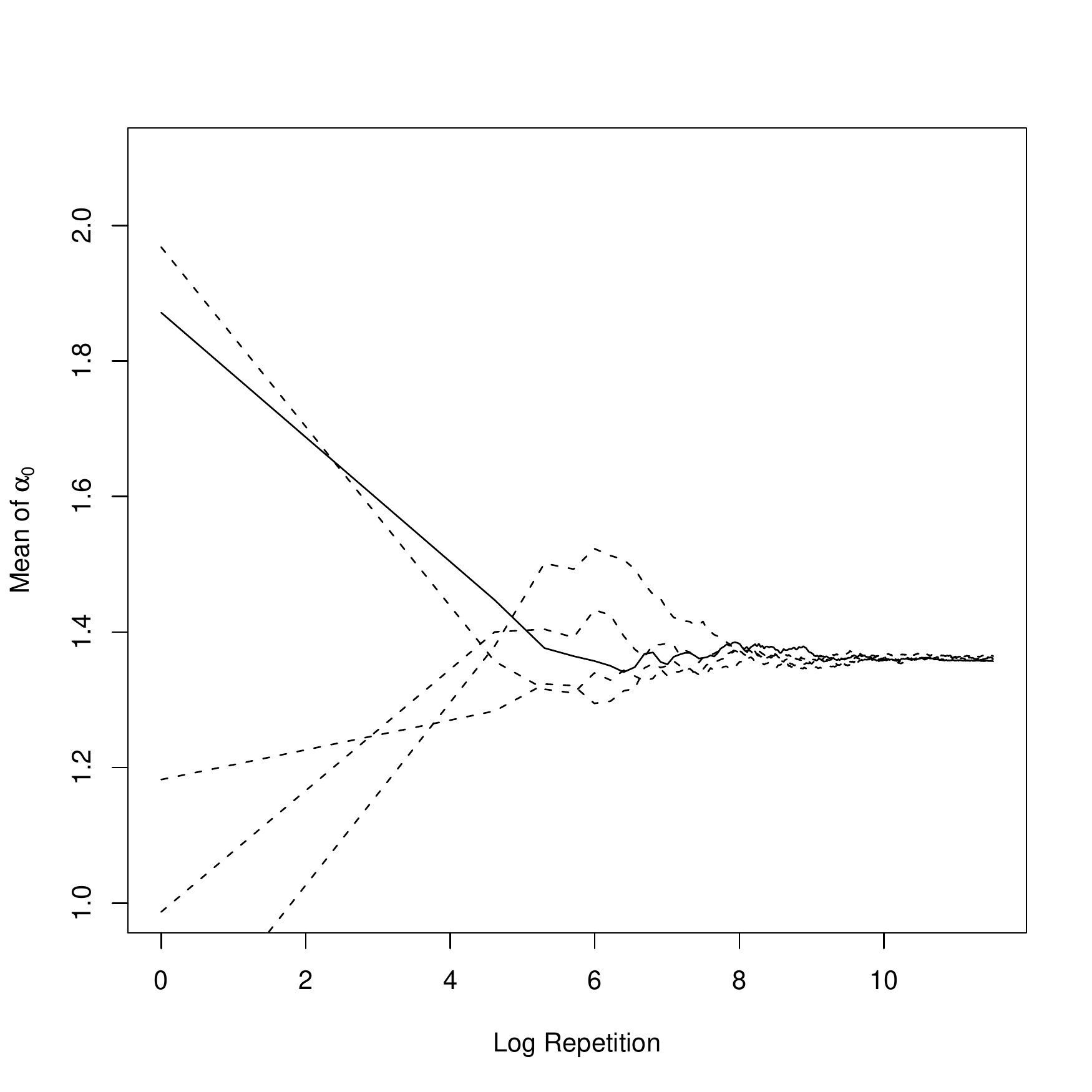}
\caption{Convergence plot for $\alpha$ and $\alpha_0$ across chains by log iteration.  This plot shows the running average of these two parameters across five separate instances of the GGM-iHMM algorithm.  Their mutual agreement implies the settings used are sufficient to assure convergence.}\label{fig:converge}
\end{figure}

\begin{figure}
\centering\includegraphics[width = 3in]{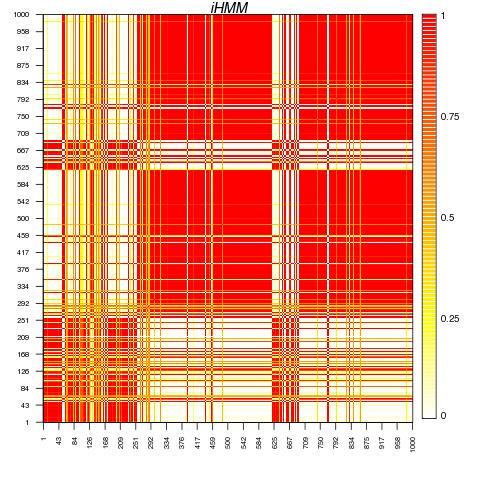}
\caption{Heatmap displaying probability that two observations belong to the same cluster for the exchange example.  The figure shows two dominant regimes, one that runs for the most part from timepoints 1 to 251 and again roughly between timepoints 623 and 700 and the other which is present most of the remaining time periods.}\label{fig:exchange_cluster}
\end{figure}

\begin{figure}[t]
\includegraphics[width = 3in]{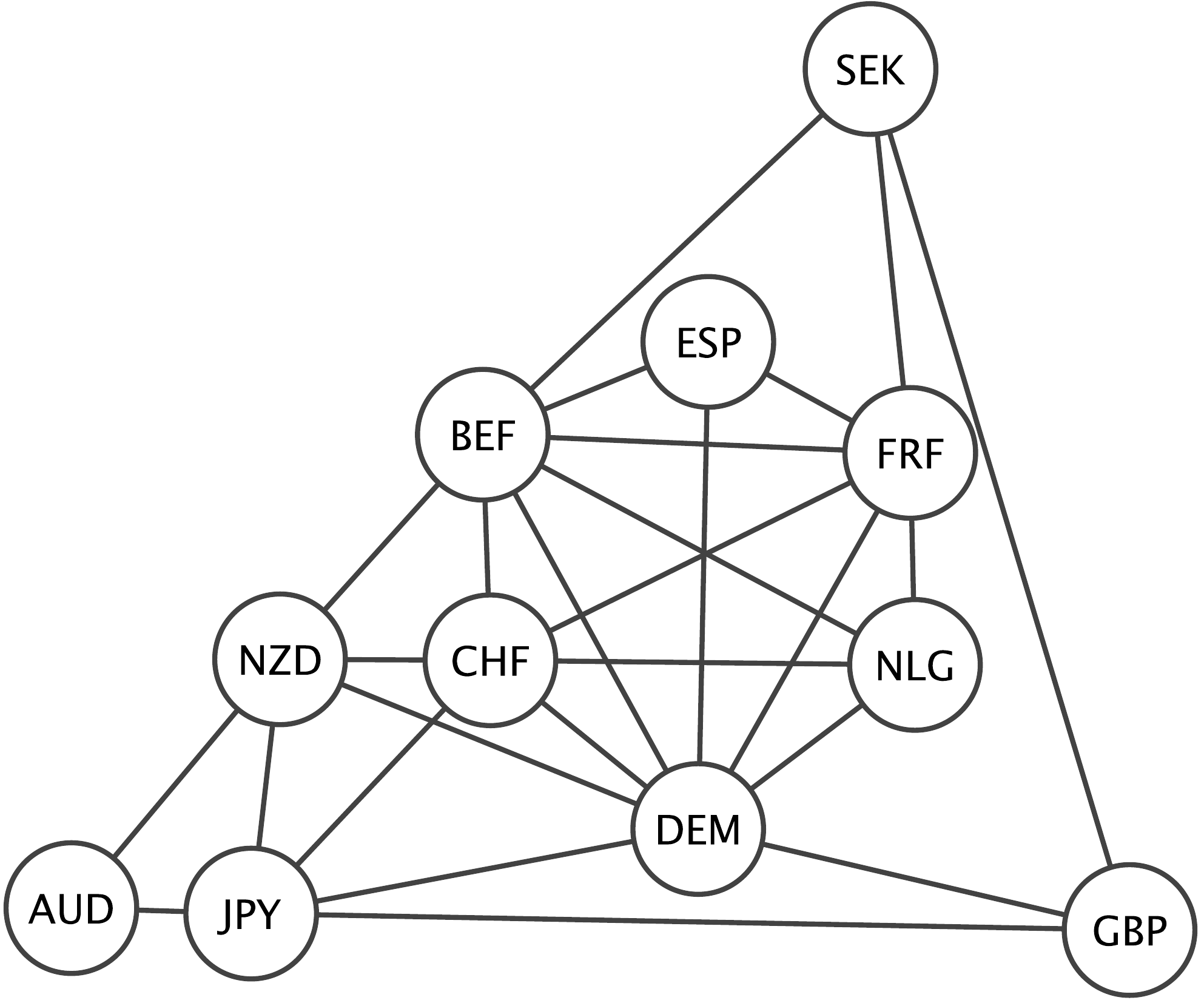}
\hspace{1cm}
\includegraphics[width = 2.35in]{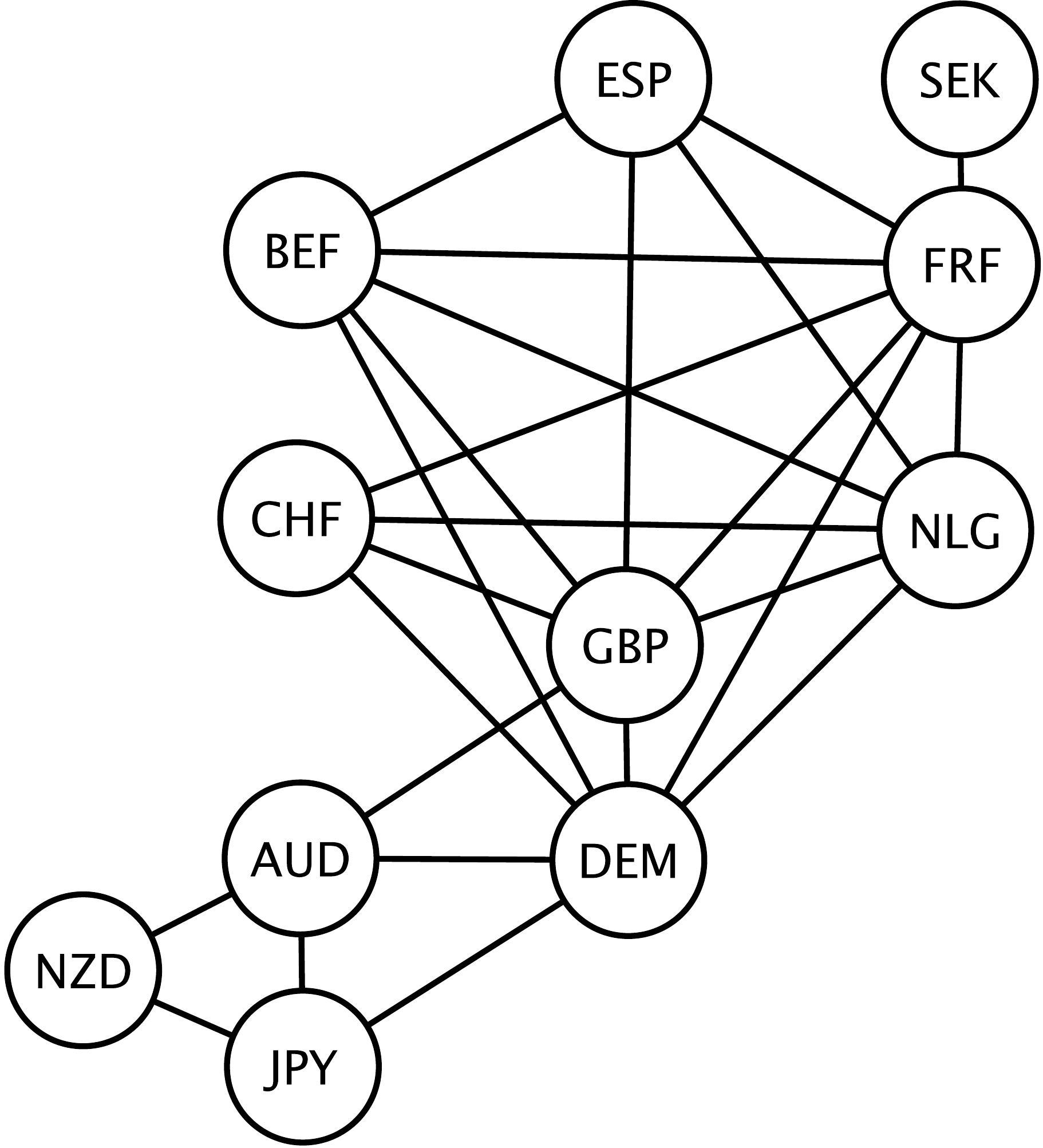}
\caption{Graphical models associated with timepoint 40 (left) and timepoint 540 (right) in the exchange rate example.  These graphs were constructed by adding any edge that had greater than 80\% posterior inclusion probability for the respective timepoint.  Such a high threshold was chosen in order to make a suitable comparison to the graph shown in Figure~\ref{fig:carvalho_graph}, which is the highest probability graph determined using stochastic search methods.}\label{fig:ihmm_graph}
\end{figure}

\begin{figure}
\centering\includegraphics[width = 4in]{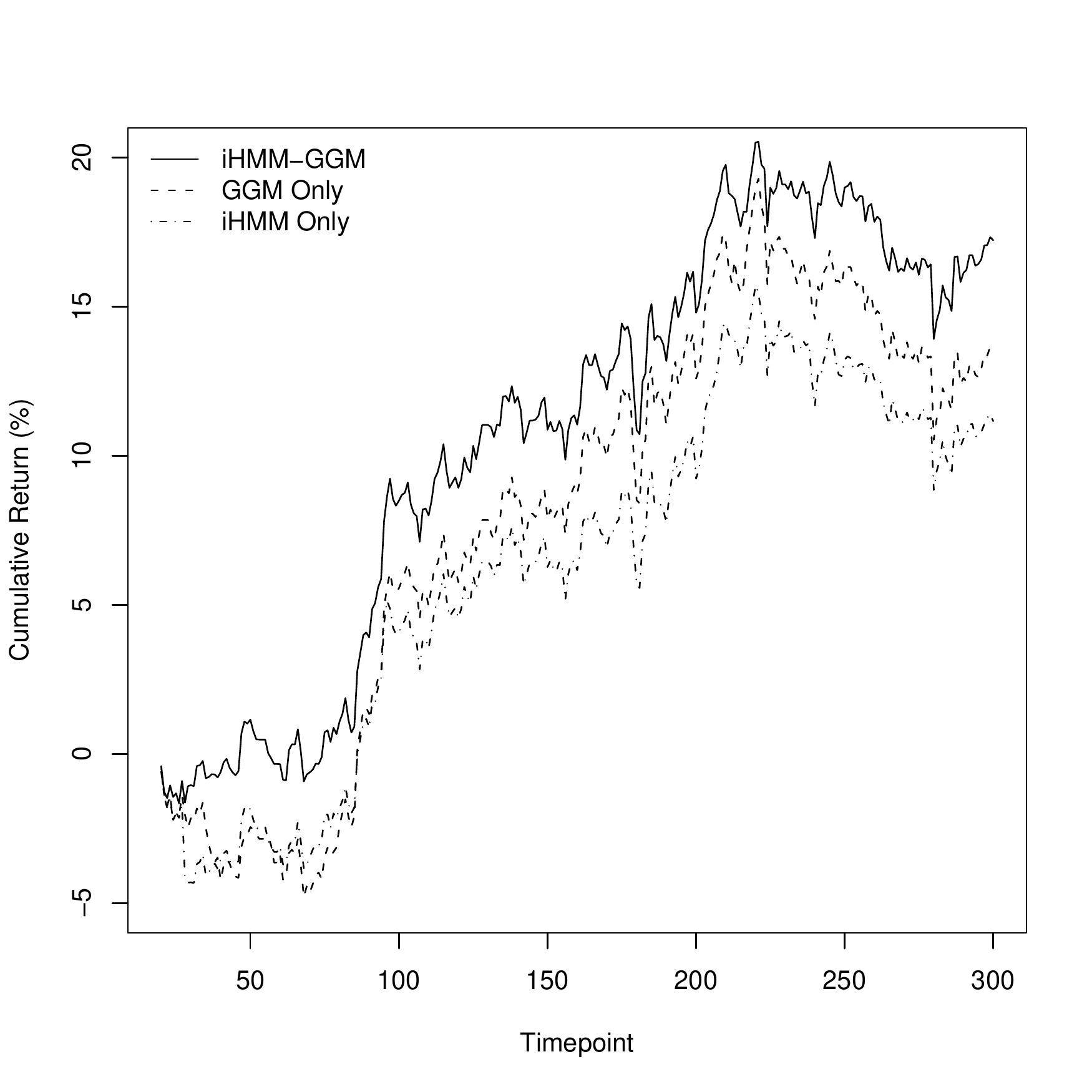}
\caption{Cumulative returns from forming the optimal portfolio weights based on running the full GGM-iHMM model as well as the iHMM only and GGM only models, run over time periods $20$ to $300$.  The final cumulative return was 17.2\% for the GGM-iHMM, 13.5\% for the GGM only and 11.2\% for the iHMM only models.}\label{fig:return}
\end{figure}

First, we consider a dataset that follows the exchange rate of 11 currencies relative to the US dollar between November 1993 and August 1996.  This dataset consists of 1000 daily observations and includes three Asian currencies (NZD,AUD, JPY), five European currencies that eventually became part of the Euro (DEM, FRF, BEF, NLG, ESP) and three additional European currencies (GBP, SEK, CHF).  These data have previously been used in a variety of contexts related to graphical models (see e.g. \citealp{carvalho_et_2007} and \citealp{CaWe07}).  In \cite{carvalho_et_2007} the authors present the graph shown in Figure~\ref{fig:carvalho_graph}, which is determined using stochastic search methods first discussed in \cite{jones_et_2005} over the final 100 timepoints.  The authors note that this graph is sensible from the standpoint of known trading relations: the mainland European countries that join the Euro are closely linked in a single clique, while the British Pound (GBP), Swedish Krona (SEK) and Swiss Franc (CHZ) connect with only some of these counties, most notably the currencies of the largest Euro-area countries, the Deutsch Mark (DEM) and French Franc (FRF) (the Swiss Franc is also connected to the Netherlands Gilder (NLG), being more integrated with mainland European economies).  \cite{CaWe07} then show that portfolio weights based on estimates from this graphical model give an investment strategy with increased return and reduced variability when compared to using an approach that does not impose graphical structure on the estimates of $\Sigma$, evidence of the effectiveness of the graphical models approach.\\
\indent We used the exchange dataset to explore the possibility of alternating regimes with separate patterns of interaction during these 1000 days.  Given that the data form a natural timecourse, we employed the GGM-iHMM model discussed in Section \ref{se:ihmm-ggm}.  We ran the model for 100000 iterations after a burn-in period of 20000 iterations, and ran five separate instances of the algorithm from separate starting points.  After completion we assessed the results from each chain and verified they returned the same estimates, Figure~\ref{fig:converge} shows the convergence in $\alpha$ and $\alpha_0$ across chains.  When run using the iHMM model, the observations for the most part clearly fall into one of two regimes.  Figure~\ref{fig:exchange_cluster} shows the posterior probabilities that two observations belong to the same state.  The first state is comprised roughly of the time points one through 251 (11/14/1993 to 7/21/1994) at which point a second regime takes over.  Interestingly the first interaction regime arises again for a brief period roughly comprising timepoints 623 to 700 (7/29/95 to 10/14/95). These two regimes show a good deal of similarity in the associated graphical models, but also present some key differences.  Figure~\ref{fig:ihmm_graph} displays the graphical model associated with timepoint 40 (belonging to the first regime) and timepoint 540 (belonging to the second).  The graphs shown are assembled from those edges that had a greater than 80\% chance of inclusion for the respective observation, which is similar in spirit to displaying the top graphical model from a stochastic search, as performed by  \cite{carvalho_et_2007}.\\
\indent In the first regime, an association structure broadly consistent with the graph used in Carvalho and West \cite{CaWe07} is present.  We see a tight grouping of the Euro adopters (not quite a clique, but missing only an ESP NLG connection), but with increased connection between the Swiss Franc and the Euro countries. A clique amongst the Asian currencies is connnected to only three of the Euro adopters.  Furthermore the Pound is only connected to the Euro through DEM, the currency of the economic leader of this area.  The interpretation of this graph is similar to that reported earlier: the fluctuations in the exchange rate of mainland European currencies to the dollar roughly track one another.  However, at this point the British Pound was no longer part of the European Exchange Rate Mechanism (ERM), following the Pound's crash on ``Black Wednesday'', September 16th of 1992.  This reason, along with the greater integration of trade between Britain and the US (as suggested by Carvalho et al \citealp{carvalho_et_2007}), leads to a separation of the Pound from the mainland European currencies.\\
 \indent The second regime has a similar structure but a markedly different interpretation of the interactions between the Pound and the smaller Euro adopters.  At this point, the graph still is comprised of a large group consisting of the Euro countries, however the Pound has joined--and become a central part of--this grouping.  It connects with each member of the Euro countries (as well as CHF) and subsequently connections between ESP and DEM and BEF and CHF are no longer present.  Furthermore, the Asian countries lose two neighbors, CHF and BEF, and the AUD is now the sole currency connecting the Asian countries to both the DEM and GBP, though the JPY maintains its association with DEM.\\
\indent The greater connectedness of the GBP to the euro area may have been a result of the uncertainty regarding the specifics of the Euro's implementation in the mid-nineties.  In particular, the crash of the Pound in 1992 and Britain's subsequent withdrawal from the ERM left a looming uncertainty regarding if, and when, Britain would again agree to join the common currency.  The initial switch from a ``UK-excluding'' to a ``UK-inclusive'' regime in the exchange rate data occurs on July 22, 1994.  What is curious about this date is that Tony Blair was elected to lead the Labour party on July 21, 1994.  Blair would eventually run a campaign based, in part, on rejoining the ERM and adopting the Euro, a stance he held until the events of 2001.  The graphical models displayed in Figure~\ref{fig:ihmm_graph} suggest that currency markets began integrating the possibility that the Pound would join the Euro by exhibiting greater covariation with mainland European countries.  This new regime was, itself, somewhat unstable, as evidenced by the return of a ``UK-exclusive'' regime during the summer of 1995.\\
\indent \cite{CaWe07} use the exchange rate data to show that minimum variance portfolios will yield better return when GGMs are employed to estimate covariances.  We considered a similar analysis and used the expectation over the sampled values of $\mu_{T + 1}$ and $K_{T + 1}$ for each $T$ between $20$ and $300$ as the first two moments of the predictive distribution and calculated portfolio weights $w_{T}$ assuming a target return of $m = 0.1\%$ per day (see \cite{CaWe07} for details regarding the construction of these portfolios).  Figure~\ref{fig:return} shows that the predictive distributions from the GGM-iHMM model give portfolio weights that have higher yields than both the iHMM only model and the GGM only model.  This shows the ulitility of our framework: by considering a mixture model we are able to adapt to changing conditions, thereby leading to better specified predictive distributions.  Furthermore, by incorporating the GGMs into the model formulation, we are able to induce sparser estimates of covariation, which likewise improve predictive performance.\\

\section{Discussion}\label{se:discussion}

Although this paper has focused on two relatively simple models (nonparametric mixtures of GGMs and infinite hidden Markov models with GGM emission distributions), the basic structure can be employed to generalize many other nonparametric models.  For example, we plan to extend the nonparametric mixture classifier developed in \cite{RoVu09} to include GGM kernels as a way to improve classification rates in high-dimensional problems.  Also, in the spirit of \cite{MuErWe96}, \cite{MuQuRo04} and \cite{RoDuGe08b}, sparse nonlinear regression models can be generated by using GGM mixtures as the joint model for outcomes and predictors, from which the regression function can be derived by computing the conditional expectation of the outcome given the predictors.  This generalizes the work of \cite{DoEiLe08} on sparse regression to allow for adaptive local linear fits.

The implementation of GGM mixtures we have discussed in this paper exploits the P{\'o}lya urn representation available for many nonparametric models to construct a Gibbs sampler that updates the grouping structure one observation at a time.  This has allowed us to avoid the explicit representation (and the sampling) of means and covariance parameters, which can be computationally intensive.  However, P{\'o}lya urn samplers can suffer from slow mixing and we plan to explore in the near future alternative computational algorithms, in particular those employing split-merge moves such as those developed in \cite{JaNe04} and \cite{JaNe07}.

The data analyses in this paper suggest that implementing mixtures of GGMs using Markov chain Monte Carlo algorithms is feasible for a moderate numbers of variables.  However, many interesting applications of GGM mixtures (e.g., gene-expression data) involve outcomes in much higher dimensions, where previous experience suggest that random-walk MCMC algorithms will be inefficient.  We are currently exploring deterministic search algorithms based on heuristics, such as the feature-inclusion \citep{BeMo05,scott_carvalho_2008}, that might allow us to identify high-probability partitions and their associated graphs.  

Our framework extends to include nondecomposable graphs in a straightforward manner. The only difference comes in the implementation of our sampling algorithms: for nondecomposable graphs the marginal likelihood (\ref{eq:marglik}) and the predictive distribution (\ref{eq:preddistrib}) must be numerically approximated instead of being calculated directly through formulas. We are currently working on this extension based on the recent developments from \citet{lenkoski-dobra-2010}.


\section*{Acknowledgments}

We would like to thank Mike West for providing access to the currency data set.  AR was partially supported by grant NSF-DMS 0915272.

\section*{Appendix}

We give a brief description of an auxiliary variable scheme for sampling from the posterior distributions of the single concentration parameter $\alpha_0$ in Section \ref{se:compu} and the two concentration parameters $\alpha$ and $\alpha_{0}$ from Section \ref{se:ihmm-ggm} \--- see \cite{EsWe95} and \cite{TeJoBeBl04} for full details. We assume that the priors for $\alpha$ and $\alpha_{0}$ are $\Gam(a,b)$ and $\Gam(a_{0},b_{0})$, respectively.  

In the case of the GGM-DPM, we sample $\alpha_{0}$ by introducing an auxiliary variable $\eta$. Conditional on $\alpha_{0}$, we have $\eta | \alpha_0 \sim \bet(\alpha_{0}+1, n)$. Conditional on $\eta$, $\alpha_{0}$ follows a mixture distribution
\begin{eqnarray*}
\alpha_{0} | \eta \sim d_{\eta}\Gam(a_{0}+L,b_{0}-\log \eta) + {(1-d_{\eta})} \Gam(a_{0}+L-1,b_{0}-\log \eta),
\end{eqnarray*}
where $d_{\eta}/(1-d_{\eta}) = (a_{0} + L-1) / [n(b_{0} - \log(\eta))]$.

In the case of the GGM-iHMM we additionally introduce auxiliary variables $\varsigma_1,\ldots,\varsigma_L$ and $u_1,\ldots,u_L$.  Conditionally of $\alpha$, $\varsigma_l | \alpha \sim \bet( \alpha + 1, r_{l \cdot})$ and $u_l | \alpha \sim \bern( r_{l \cdot} / (\alpha + r_{l\cdot}))$, where $r_{l\cdot} = \sum_{l'=1}^{L} r_{ll'}$.  Then, $\alpha$ is sampled from its full conditional distribution,
$$
\alpha | \{ u_l \}, \{ \varsigma_l \} \sim \Gam\left( a + m_{\cdot \cdot} - \sum_{l=1}^{L} u_{l}, b - \sum_{l=1}^{L} \log \varsigma_l \right)
$$
where $m_{\cdot\cdot} = \sum_{l=1}^{L} \sum_{l'=1}^{L} m_{l l'}$.  To sample $\alpha_0$, we follow a procedure that is very similar to the one we used for the GGM-DPM.  Again, we introduce an auxiliary variable $\eta$. Conditional on $\alpha_{0}$, we have $\eta | \alpha_0 \sim \bet(\alpha_{0}+1, m_{\cdot \cdot})$. Conditional on $\eta$, $\alpha_{0}$ follows a mixture distribution
\begin{eqnarray*}
\alpha_{0} | \eta \sim d_{\eta}\Gam(a_{0}+L,b_{0}-\log \eta) + {(1-d_{\eta})} \Gam(a_{0}+L-1,b_{0}-\log \eta),
\end{eqnarray*}
where $d_{\eta}/(1-d_{\eta}) = (a_{0} + L-1) / [m_{\cdot \cdot}(b_{0} - \log(\eta))]$.

\bibliographystyle{bka}
\bibliography{ggm-01232010}

\begin{thebibliography}{61}
\providecommand{\natexlab}[1]{#1}
\expandafter\ifx\csname urlstyle\endcsname\relax
  \providecommand{\doi}[1]{doi:\discretionary{}{}{}#1}\else
  \providecommand{\doi}{doi:\discretionary{}{}{}\begingroup
  \urlstyle{rm}\Url}\fi

\bibitem[{Antoniak(1974)}]{An74}
\textsc{Antoniak, C.} (1974).
\newblock Mixtures of {D}irichlet processes with applications to {B}ayesian
  nonparametric problems.
\newblock \emph{Annals of Statistics} \textbf{{\bf 2}}, 1152--1174.

\bibitem[{Atay-Kayis \& Massam(2005)}]{atay-kayis_massam_2005}
\textsc{Atay-Kayis, A.} \& \textsc{Massam, H.} (2005).
\newblock A {M}onte {C}arlo method for computing the marginal likelihood in
  nondecomposable {G}aussian graphical models.
\newblock \emph{Biometrika} \textbf{92}, 317--35.

\bibitem[{Beal et~al.(2001)Beal, Ghahramani \& Rasmussen}]{BeGhRa01}
\textsc{Beal, M.~J.}, \textsc{Ghahramani, Z.} \& \textsc{Rasmussen, C.~E.}
  (2001).
\newblock The infinite hidden markov model.
\newblock In \emph{Proceedings of Fourteenth Annual Conference on Neural
  Information Processing Systems}.

\bibitem[{Bedford \& Cooke(2002)}]{BeCo02}
\textsc{Bedford, T.} \& \textsc{Cooke, R.~M.} (2002).
\newblock Vines - a new graphical model for dependent random variables.
\newblock \emph{Annals of Statistics} \textbf{{\bf 30}}, 1031--1068.

\bibitem[{Berger \& Molina(2005)}]{BeMo05}
\textsc{Berger, J.~O.} \& \textsc{Molina, G.} (2005).
\newblock Posterior model probabilities via path-based pairwise priors.
\newblock \emph{Statistica Neerlandica} \textbf{{\bf 59}}, 3--15.

\bibitem[{Blackwell \& MacQueen(1973)}]{BlMQ73}
\textsc{Blackwell, D.} \& \textsc{MacQueen, J.~B.} (1973).
\newblock Ferguson distribution via {P}{\'o}lya urn schemes.
\newblock \emph{The Annals of Statistics} \textbf{{\bf 1}}, 353--355.

\bibitem[{Capp{\'e} et~al.(2005)Capp{\'e}, Moulines \& Ryden}]{CaMoRy05}
\textsc{Capp{\'e}, O.}, \textsc{Moulines, E.} \& \textsc{Ryden, T.} (2005).
\newblock \emph{Inference in Hidden Markov Models}.
\newblock Springer.

\bibitem[{Carvalho et~al.(2007)Carvalho, Massam \& West}]{carvalho_et_2007}
\textsc{Carvalho, C.~M.}, \textsc{Massam, H.} \& \textsc{West, M.} (2007).
\newblock Simulation of hyper-inverse {W}ishart distributions in graphical
  models.
\newblock \emph{Biometrika} \textbf{{\bf 94}}, 647--659.

\bibitem[{Carvalho \& Scott(2009)}]{carvalho_scott_2009}
\textsc{Carvalho, C.~M.} \& \textsc{Scott, J.~G.} (2009).
\newblock Objective {B}ayesian model selection in {G}aussian graphical models.
\newblock \emph{Biometrika} To appear.

\bibitem[{Carvalho \& West(2007)}]{CaWe07}
\textsc{Carvalho, C.~M.} \& \textsc{West, M.} (2007).
\newblock Dynamic matrix-variate graphical models.
\newblock \emph{Bayesian Analysis} \textbf{{\bf 2}}, 69--98.

\bibitem[{Castelo \& Roverato(2006)}]{CaRo06}
\textsc{Castelo, R.} \& \textsc{Roverato, A.} (2006).
\newblock A robust procedure for {G}aussian graphical model search from
  microarray data with p larger than n.
\newblock \emph{J. Mach. Learn. Res.} \textbf{7}, 2621--2650.

\bibitem[{Dawid \& Lauritzen(1993)}]{dawid-lauritzen-1993}
\textsc{Dawid, A.~P.} \& \textsc{Lauritzen, S.~L.} (1993).
\newblock Hyper {M}arkov laws in the statistical analysis of decomposable
  graphical models.
\newblock \emph{Ann. Statist.} \textbf{21}, 1272--1317.

\bibitem[{Dempster(1972)}]{dempster_1972}
\textsc{Dempster, A.~P.} (1972).
\newblock Covariance selection.
\newblock \emph{Biometrics} \textbf{28}, 157--75.

\bibitem[{Diaconnis \& Ylvisaker(1979)}]{diaconis_ylvisaker_1979}
\textsc{Diaconnis, P.} \& \textsc{Ylvisaker, D.} (1979).
\newblock Conjugate priors for exponential families.
\newblock \emph{Ann. Statist.} \textbf{7}, 269--81.

\bibitem[{Dobra et~al.(2008)Dobra, Eicher \& Lenkoski}]{DoEiLe08}
\textsc{Dobra, A.}, \textsc{Eicher, T.~S.} \& \textsc{Lenkoski, A.} (2008).
\newblock Modeling uncertainty in macroeconomic growth determinants using
  {G}aussian graphical models.
\newblock Technical Report~55, Center for Statistics and the Social Sciences,
  University of Washington.

\bibitem[{Dobra et~al.(2004)Dobra, Hans, Jones, Nevins, Yao \&
  West}]{DoHaJoNeYaWe04}
\textsc{Dobra, A.}, \textsc{Hans, C.}, \textsc{Jones, B.}, \textsc{Nevins,
  J.~R.}, \textsc{Yao, G.} \& \textsc{West, M.} (2004).
\newblock Sparse graphical models for exploring gene expression data.
\newblock \emph{Journal of Multivariate Analysis} \textbf{{\bf 90}}, 196--212.

\bibitem[{Dunson et~al.(2007)Dunson, Pillai \& Park}]{DuPi04}
\textsc{Dunson, D.~B.}, \textsc{Pillai, N.} \& \textsc{Park, J.-H.} (2007).
\newblock {B}ayesian density regression.
\newblock \emph{Journal of the Royal Statistical Society, Series B.}
  \textbf{{\bf 69}}, 163--183.

\bibitem[{Escobar \& West(1995)}]{EsWe95}
\textsc{Escobar, M.~D.} \& \textsc{West, M.} (1995).
\newblock {B}ayesian density estimation and inference using mixtures.
\newblock \emph{Journal of American Statistical Association} \textbf{{\bf 90}},
  577--588.

\bibitem[{Ferguson(1973)}]{Fe73}
\textsc{Ferguson, T.~S.} (1973).
\newblock A {B}ayesian analysis of some nonparametric problems.
\newblock \emph{Annals of Statistics} \textbf{{\bf 1}}, 209--230.

\bibitem[{Ferguson(1974)}]{Fe74}
\textsc{Ferguson, T.~S.} (1974).
\newblock Prior distributions on spaces of probability measures.
\newblock \emph{Annals of Statistics} \textbf{{\bf 2 }}, 615--629.

\bibitem[{Fraley \& Raftery(2007)}]{FrRa07}
\textsc{Fraley, C.} \& \textsc{Raftery, A.~E.} (2007).
\newblock {B}ayesian regularization for normal mixture estimation and
  model-based clustering.
\newblock \emph{Journal of Classification} \textbf{{\bf 24}}, 155--181.

\bibitem[{Friedman(2004)}]{Fr04}
\textsc{Friedman, N.} (2004).
\newblock Inferring cellular networks using probabilistic graphical models.
\newblock \emph{Science} \textbf{{\bf 6}}, 799--805.

\bibitem[{van Gael et~al.(2008)van Gael, Saatci, Teh \&
  Ghahramani}]{GaSaTeGh08}
\textsc{van Gael, J.}, \textsc{Saatci, Y.}, \textsc{Teh, Y.-W.} \&
  \textsc{Ghahramani, Z.} (2008).
\newblock Beam sampling for the infinite hidden markov model.
\newblock In \emph{Proceedings of the 25th International Conference on Machine
  Learning (ICML)}.

\bibitem[{Green \& Richardson(2001)}]{GrRi01}
\textsc{Green, P.} \& \textsc{Richardson, S.} (2001).
\newblock Modelling heterogeneity with and without the {D}irichlet process.
\newblock \emph{Scandinavian Journal of Statistics} \textbf{{\bf 28}},
  355--375.

\bibitem[{Green(1995)}]{Gr95}
\textsc{Green, P.~J.} (1995).
\newblock Reversible jump {M}arkov chain {M}onte {C}arlo computation and
  {B}ayesian model determination.
\newblock \emph{Biometrika} \textbf{{\bf 82}}.

\bibitem[{Ishwaran \& James(2001)}]{IsJa01}
\textsc{Ishwaran, H.} \& \textsc{James, L.~F.} (2001).
\newblock {G}ibbs sampling methods for stick-breaking priors.
\newblock \emph{Journal of the American Statistical Association} \textbf{{\bf
  96}}, 161--173.

\bibitem[{Ishwaran \& Zarepour(2002)}]{IsZa02}
\textsc{Ishwaran, H.} \& \textsc{Zarepour, M.} (2002).
\newblock {D}irichlet prior sieves in finite normal mixtures.
\newblock \emph{Statistica Sinica} \textbf{{\bf 12}}, 941--963.

\bibitem[{Jain \& Neal(2004)}]{JaNe04}
\textsc{Jain, S.} \& \textsc{Neal, R.~M.} (2004).
\newblock A split-merge {M}arkov chain {M}onte {C}arlo procedure for the
  {D}irichlet process mixture model.
\newblock \emph{Journal of Graphical and Computational Statistics} \textbf{{\bf
  13}}, 158--182.

\bibitem[{Jain \& Neal(2007)}]{JaNe07}
\textsc{Jain, S.} \& \textsc{Neal, R.~M.} (2007).
\newblock Splitting and merging components of a nonconjugate dirichlet process
  mixture model.
\newblock \emph{Bayesian Analysis} \textbf{{\bf 2}}, 445--472.

\bibitem[{Jones et~al.(2005)Jones, Carvalho, Dobra, Hans, Carter \&
  West}]{jones_et_2005}
\textsc{Jones, B.}, \textsc{Carvalho, C.}, \textsc{Dobra, A.}, \textsc{Hans,
  C.}, \textsc{Carter, C.} \& \textsc{West, M.} (2005).
\newblock Experiments in stochastic computation for high-dimensional graphical
  models.
\newblock \emph{Statist. Sci.} \textbf{20}, 388--400.

\bibitem[{Lauritzen(1996)}]{lauritzen_1996}
\textsc{Lauritzen, S.~L.} (1996).
\newblock \emph{Graphical Models}.
\newblock Oxford University Press.

\bibitem[{Lee et~al.(2009)Lee, M\"{u}ller, Trippa \& Quintana}]{LeMuTrQu09}
\textsc{Lee, J.}, \textsc{M\"{u}ller, P.}, \textsc{Trippa, L.} \&
  \textsc{Quintana, F.~A.} (2009).
\newblock Defining predictive probability functions for species sampling
  models.
\newblock Technical report, Pontificia Universidad Cat{\'o}lica de Chile.

\bibitem[{Lenkoski \& Dobra(2010)}]{lenkoski-dobra-2010}
\textsc{Lenkoski, A.} \& \textsc{Dobra, A.} (2010).
\newblock Computational aspects related to inference in {G}aussian graphical
  models with the {G}-wishart prior.
\newblock \emph{Journal of Computational and Graphical Statistics} To appear.

\bibitem[{Letac \& Massam(2007)}]{letac_massam_2007}
\textsc{Letac, G.} \& \textsc{Massam, H.} (2007).
\newblock {W}ishart distributions for decomposable graphs.
\newblock \emph{Ann. Statist.} \textbf{35}, 1278--323.

\bibitem[{Lijoi et~al.(2005)Lijoi, Mena \& Pr{\"u}nster}]{LiMePr05b}
\textsc{Lijoi, A.}, \textsc{Mena, R.~H.} \& \textsc{Pr{\"u}nster, I.} (2005).
\newblock Hierarchical mixture modelling with normalized inverse {G}aussian
  priors.
\newblock \emph{Journal of American Statistical Association} \textbf{{\bf
  100}}, 1278--1291.

\bibitem[{Lijoi et~al.(2007)Lijoi, Mena \& Pr{\"u}nster}]{LiMePr07}
\textsc{Lijoi, A.}, \textsc{Mena, R.~H.} \& \textsc{Pr{\"u}nster, I.} (2007).
\newblock Bayesian nonparametric estimation of the probability of discovering
  new species.
\newblock \emph{Biometrika} \textbf{{\bf 94}}, 769--786.

\bibitem[{Lo(1984)}]{Lo84}
\textsc{Lo, A.~Y.} (1984).
\newblock On a class of {B}ayesian nonparametric estimates: I. {D}ensity
  estimates.
\newblock \emph{Annals of Statistics} \textbf{{\bf 12}}, 351--357.

\bibitem[{MacEachern(1994)}]{ME94}
\textsc{MacEachern, S.~N.} (1994).
\newblock Estimating normal means with a conjugate style {D}irichlet process
  prior.
\newblock \emph{Commnunications in Statistics, Part B - Simulation and
  Computation} \textbf{{\bf 23}}, 727--741.

\bibitem[{McCloskey(1965)}]{MC65}
\textsc{McCloskey, J.~W.} (1965).
\newblock \emph{A Model for the Distribution of Individuals by Species in an
  Environment}.
\newblock Ph.D. thesis, Michigan State University.

\bibitem[{Muirhead(2005)}]{muirhead_2005}
\textsc{Muirhead, R.~J.} (2005).
\newblock \emph{Aspects of Multivariate Statistical Theory}.
\newblock John Wiley \& Sons.

\bibitem[{M\"{u}ller et~al.(1996)M\"{u}ller, Erkanli \& West}]{MuErWe96}
\textsc{M\"{u}ller, P.}, \textsc{Erkanli, A.} \& \textsc{West, M.} (1996).
\newblock {B}ayesian curve fitting using multivariate normal mixtures.
\newblock \emph{Biometrika} \textbf{{\bf 83}}, 67--79.

\bibitem[{M\"{u}ller et~al.(2004)M\"{u}ller, Quintana \& Rosner}]{MuQuRo04}
\textsc{M\"{u}ller, P.}, \textsc{Quintana, F.} \& \textsc{Rosner, G.} (2004).
\newblock Hierarchical meta-analysis over related non-parametric {B}ayesian
  models.
\newblock \emph{Journal of Royal Statistical Society, Series B} \textbf{{\bf
  66}}, 735--749.

\bibitem[{Neal(2000)}]{Ne00}
\textsc{Neal, R.~M.} (2000).
\newblock {M}arkov chain sampling methods for {D}irichlet process mixture
  models.
\newblock \emph{Journal of Computational and Graphical Statistics} \textbf{{\bf
  9}}, 249--.

\bibitem[{Ongaro \& Cattaneo(2004)}]{OnCa04}
\textsc{Ongaro, A.} \& \textsc{Cattaneo, C.} (2004).
\newblock Discrete random probability measures: a general framework for
  nonparametric {B}ayesian inference.
\newblock \emph{Statistics and Probability Letters} \textbf{{\bf 67}}, 33--45.

\bibitem[{Pitman(1996)}]{Pi96}
\textsc{Pitman, J.} (1996).
\newblock Some developments of the blackwell-macqueen urn scheme.
\newblock In \emph{Statistics, Probability and Game Theory. Papers in Honor of
  David Blackwell}, Eds. T.~S. Ferguson, L.~S. Shapeley \& J.~B. MacQueen, pp.
  245--268. Hayward, CA:IMS.

\bibitem[{Roberts \& Papaspiliopoulos(2008)}]{RoPa04}
\textsc{Roberts, G.} \& \textsc{Papaspiliopoulos, O.} (2008).
\newblock Retrospective {M}arkov chain {M}onte {C}arlo methods for {D}irichlet
  process hierarchical models.
\newblock \emph{Biometrika} \textbf{{\bf 95}}, 169--186.

\bibitem[{Rodriguez et~al.(2008)Rodriguez, Dunson \& Gelfand}]{RoDuGe08a}
\textsc{Rodriguez, A.}, \textsc{Dunson, D.~B.} \& \textsc{Gelfand, A.~E.}
  (2008).
\newblock The nested {D}irichlet process, with discussion.
\newblock \emph{Journal of American Statistical Association} \textbf{{\bf
  103}}, 1131--1144.

\bibitem[{Rodriguez et~al.(2009)Rodriguez, Dunson \& Gelfand}]{RoDuGe08b}
\textsc{Rodriguez, A.}, \textsc{Dunson, D.~B.} \& \textsc{Gelfand, A.~E.}
  (2009).
\newblock Bayesian nonparametric functional data analysis through density
  estimation.
\newblock \emph{Biometrika} \textbf{{\bf 96}}, 149--162.

\bibitem[{Rodriguez \& Vuppala(2009)}]{RoVu09}
\textsc{Rodriguez, A.} \& \textsc{Vuppala, R.} (2009).
\newblock Probabilistic classification using bayesian nonparametric mixture
  models.
\newblock Technical report, University of California, Santa Cruz.

\bibitem[{Roverato(2002)}]{roverato_2002}
\textsc{Roverato, A.} (2002).
\newblock Hyper inverse {W}ishart distribution for non-decomposable graphs and
  its application to {B}ayesian inference for {G}aussian graphical models.
\newblock \emph{Scand. J. Statist.} \textbf{29}, 391--411.

\bibitem[{Scott \& Berger(2006)}]{scott_berger_2006}
\textsc{Scott, J.~G.} \& \textsc{Berger, J.~O.} (2006).
\newblock An exploration of aspects of {B}ayesian multiple testing.
\newblock \emph{J. Statist. Plan. Infer.} \textbf{136}, 2144--2162.

\bibitem[{Scott \& Carvalho(2008)}]{scott_carvalho_2008}
\textsc{Scott, J.~G.} \& \textsc{Carvalho, C.~M.} (2008).
\newblock Feature-inclusion stochastic search for {G}aussian graphical models.
\newblock \emph{J. Comput. Graph. Statist.} \textbf{17}, 790--808.

\bibitem[{Sethuraman(1994)}]{Se94}
\textsc{Sethuraman, J.} (1994).
\newblock A constructive definition of {D}irichelt priors.
\newblock \emph{Statistica Sinica} \textbf{{\bf 4}}, 639--650.

\bibitem[{Sudderth \& Jordan(2009)}]{SuJo09}
\textsc{Sudderth, E.~B.} \& \textsc{Jordan, M.~I.} (2009).
\newblock Shared segmentation of natural scenes using dependent {P}itman-{Y}or
  processes.
\newblock In \emph{Advances in Neural Information Processing Systems 21}, Eds.
  D.~Koller, D.~Schuurmans, Y.~Bengio \& L.~Bottou.

\bibitem[{Teh et~al.(2006)Teh, Jordan, Beal \& Blei}]{TeJoBeBl04}
\textsc{Teh, Y.~W.}, \textsc{Jordan, M.~I.}, \textsc{Beal, M.~J.} \&
  \textsc{Blei, D.~M.} (2006).
\newblock Sharing clusters among related groups: Hierarchical {D}irichlet
  processes.
\newblock \emph{Journal of the American Statistical Association} \textbf{{\bf
  101}}, 1566--1581.

\bibitem[{Thiesson et~al.(1997)Thiesson, Meek, Chickering \&
  Heckerman}]{ThMeChHe99}
\textsc{Thiesson, B.}, \textsc{Meek, C.}, \textsc{Chickering, D.~M.} \&
  \textsc{Heckerman, D.} (1997).
\newblock Learning mixtures of {DAG} models.
\newblock In \emph{Proc. of the Conf. on Uncertainty in AI}, pp. 504--513.
  Morgan Kaufmann, Inc.

\bibitem[{Wainwright et~al.(2006)Wainwright, Ravikumar \&
  Lafferty}]{Wainwright06high-dimensionalgraphical}
\textsc{Wainwright, M.~J.}, \textsc{Ravikumar, P.} \& \textsc{Lafferty, J.~D.}
  (2006).
\newblock High-dimensional graphical model selection using ℓ1-regularized
  logistic regression.
\newblock In \emph{In Neural Information Processing Systems}. MIT Press.

\bibitem[{Walker(2007)}]{Wa07}
\textsc{Walker, S.~G.} (2007).
\newblock Sampling the dirichlet mixture model with slices.
\newblock \emph{Communications in Statistics - Simulation and Computation}
  \textbf{{\bf 36}}, 45--54.

\bibitem[{Wang \& West(2009)}]{wang_west_2009}
\textsc{Wang, H.} \& \textsc{West, M.} (2009).
\newblock Bayesian analysis of matrix normal graphical models.
\newblock \emph{Biometrika} To appear.

\bibitem[{West et~al.(2001)West, Blanchette, Dressman, Huang, Ishida, Spang,
  Zuzan, Olson, Marks \& Nevings}]{WeBlDrHuIsSpZuOlMaNe01}
\textsc{West, M.}, \textsc{Blanchette, H.}, \textsc{Dressman, H.},
  \textsc{Huang, E.}, \textsc{Ishida, S.}, \textsc{Spang, R.}, \textsc{Zuzan,
  H.}, \textsc{Olson, J.~A.}, \textsc{Marks, J.~R.} \& \textsc{Nevings, J.~R.}
  (2001).
\newblock Predicting the clinical status of human breast cancer by using gne
  expression profiles.
\newblock \emph{Proceedings of the National Academi of Sciences} \textbf{{\bf
  98}}, 11462--11467.

\bibitem[{West \& Harrison(1997)}]{WeHa97}
\textsc{West, M.} \& \textsc{Harrison, J.} (1997).
\newblock \emph{{B}ayesian Forecasting and Dynamic Models}.
\newblock Springer - Verlag, New York, second edition edition.

\end{thebibliography}

\end{document}